\documentclass{article}

\begin{document}

\title{Fractional operators and special functions. II. Legendre functions\thanks{This work was supported in part by the U. S. Department of Energy under Grant No.\ DE-FG02-95ER40896, and in part by the University of Wisconsin Graduate School with funds granted by the Wisconsin Alumni Reseach Foundation.  
}}

\author{Loyal Durand\thanks{Department of Physics, University of Wisconsin,
Madison, Wisconsin 53706. Electronic address: ldurand@theory1.hep.wisc.edu}}

\maketitle
\begin{abstract}
Most of the special functions of mathematical physics are connected with the representation of Lie groups. The action of elements $D$ of the associated Lie algebras as linear differential operators gives relations among the functions in a class, for example, their differential recurrence relations. In this paper, we apply the fractional generalizations $D^\mu$ of these operators developed in an earlier paper in the context of Lie theory to the group SO(2,1) and its conformal extension. The fractional relations give a variety of interesting relations for the associated Legendre functions. We show that the two-variable fractional operator relations lead directly to integral relations among the Legendre functions and to one- and two-variable integral representations for those functions. Some of the relations reduce to known fractional integrals for the Legendre functions when reduced to one variable. The results enlarge the understanding of many properties of the associated Legendre functions on the basis of the underlying group structure.

\end{abstract}





\section{Introduction}
\label{sec:intro}

The classical special functions (Jacobi, Gegenbauer, Legendre, Laguerre, Bessel, and Hermite functions)  are all connected with the representation of Lie groups \cite{vilenkin,miller1,talman,miller2}, or more generally, to the realization of their Lie algebras by linear differential operators $D(w,\partial_w)$ acting on functions of the variables $w$. In particular, the special functions appear as factors in the multivariable functions on which the action of the Lie algebra is realized.  Many of the properties of the special functions are easily understood in this context. For example, the differential equations for the special functions are connected with the Casimir operators of the associated groups. The actions of appropriate elements $D$ of the Lie algebra lead, when reduced to a single variable, to the standard differential recurrence relations for the functions, while the action of group elements $e^{-tD}$ can be interpreted in terms of generalized generating functions when expressed using a Taylor series expansion in the group parameter $t$. Numerous examples are given in \cite{vilenkin,talman}. 

The differential recurrence relations for the special functions are schematically of the form $DF_{\alpha,\ldots}=cF_{\alpha\pm 1,\ldots}$ where the $\alpha$'s label the realization of the Lie algebra and $D$ is a stepping operator. In a previous paper \cite{FracOpsBessel}, we defined fractional generalizations $D^\mu$ of the $D$'s in the context of Lie theory, determined their formal properties, and illustrated their usefulness in obtaining further interesting relations among the functions using the group E(2) and the Bessel functions. We showed, for example, that shifts of the index $\nu$ of a Bessel function $Z_\nu$ by an arbitrary amount $\mu$ could be effected using $D^\mu$. The resulting relations for the inverse operators $D^{-\mu}$, when reduced to a single variable, gave generalizations of known fractional integrals such as the Riemann integral
\begin{equation}
\label{sonine_frac}
x^{(\nu+\mu)/2}J_{\nu+\mu}(2\sqrt{x})= \frac{1}{\Gamma(\mu)} \int_0^{x}t^{\nu/2}J_\nu(2\sqrt{t}) (x-t)^{\mu-1}\, dt 
\end{equation}
and the Weyl integral
\begin{equation}
\label{K_nu}
x^{-(\nu-\mu)/2}K_{\nu-\mu}(2\sqrt{x}) = \frac{1}{\Gamma(\mu)}\int_x^\infty t^{-\nu} K_\nu(2\sqrt{t})(t-x)^{\mu-1},
\end{equation}
\cite{TIT}, Chap.~13. Finally, we obtained integral representations for the $Z$'s as the action of the $D^\mu$'s on appropriate input functions. While most  of the specific results had been derived historically in other ways, the introduction of the fractional operators allowed them to be unified in a group setting. 

We continue that program here for the associated Legendre functions, working in the context of the group SO(2,1) and its conformal extension. We find, in particular, fractional operators which raise of lower the order $\mu$ or degree $\nu$ of a general associated Legendre function $F_\nu^\mu$ by arbitrary amounts, and use the results to unify and extend a number of known results for those functions.

We will summarize the definitions and properties of the fractional operators in the following section, and then apply the theory to derive a number of relations for the associated Legendre functions. These include generating functions, generalizations of known fraction integral relations, and some new integral relations. With appropriate choices for the input functions, the fractional operator relations give integral representations for the associated Legendre functions, and provide group-theoretical setting for those representations. We find, in particular, interesting double-integral representations.


\section{Fractional operators}
\label{sec:frac_op}

We will suppose that we have a Lie algebra which corresponds to one of the classical Lie groups, and is realized by the action of a set of linear differential operators $D(w,\partial_w)$ in variables $w$ on an appropriate class of functions $F(w)$. The exponentials $e^{-tD}$ defined by Taylor series expansion in the group parameter $t$ are elements of the Lie group, and act on the same functions. We will suppose initially that $e^{-tD}F$ exists for all $t$, and define a Weyl-type fractional operator $D_W^\mu$ by an integral over group elements, with
\begin{equation}
\label{D_mu_Weyl}
D_W^\mu F(w) = \frac{1}{2\pi i}e^{i\pi\mu}\Gamma(\mu+1)\int_{C_W}dt\,\frac{e^{-tD(w)}}{t^{\mu+1}} F(w).
\end{equation}
The contour $C_W = (\infty,0+,\infty)$ in the complex $t$ plane runs in from infinity, circles $t=0$ in the positive sense, and runs back to infinity. To define phases, we take the integrand as cut along the positive real axis with the phase of $t$ taken as zero on the upper edge of the cut. The direction of the contour at infinity must be such that the integral converges. 

The expression above would be an identity for $D$ a positive constant. Here, however, $D(w,\partial_w)$ is an operator which acts on the functions $F(w)$, and the existence of the integral depends on the functions as well as the contour.   

Alternatively, $D_W^\mu$ can be defined without the contour integral as
\begin{equation}
\label{D_mu_Weyl2}
D_W^\mu F = \frac{1}{\Gamma(-\mu+n)}\,D^n\int_0^{\,\infty} dt \frac{e^{-tD}}{t^{\mu-n+1}}F,
\end{equation}
where ${\rm Re}\,\mu<n$ and endpoint terms are assumed to vanish in the partial integrations which connect the two expressions. 

We define a second Riemann-type fractional operator $D_R^\mu F$ by
\begin{equation}
\label{D_mu_R2}
D_R^\mu F =\frac{1}{2\pi i}e^{i\pi\mu}\Gamma(\mu+1)\int_{C_R} dt\,\frac{e^{-tD}}{t^{\mu+1}}F ,
\end{equation}
where $C_R$ is the contour $C_R=\left(x(w),0+,x(w)\right)$.  The endpoint $x(w)$ of the contour must be chosen such that $D^\mu F$ satisfies a differential equation determined by the Casimir operators of the Lie algebra. This will require that a differential expression related to $e^{-tD}F$  vanish at $t=x(w)$ for the given values of $w$ (see, for example, \S\ref{subsec:RiemannLegendre}). 

Which expression for $D^\mu$ is appropriate in a particular setting, Weyl or Riemann, will depend on $D$ and $F$. We will therefore simply denote the fractional operator as $D^\mu$ for formal purposes, and only specify the expression to be used in connection with particular applications. The key restrictions will be the existence of a finite value of the group parameter $t=x(w)$ such that a differential expression related to $e^{-xD}F$ vanishes in the Riemann case, and the convergence of the integral for $t\rightarrow\infty$ in the Weyl case.

As shown in \cite{FracOpsBessel}, the fractional operators satisfy the expected product rule for powers and commute,
\begin{equation}
\label{commutator}
D^\mu D^\nu=D^\nu D^\mu=D^{\mu+\nu},\quad [D^\mu,D^\nu]=0.
\end{equation}
The inverse of $D^\mu$ is just $D^{-\mu}$,
\begin{equation}
\label{Dinverse}
(D^\mu)^{-1}=D^{-\mu},\qquad D^{-\mu}D^\mu={\bf 1}.
\end{equation}
%


\section{Legendre functions, SO(3), and SO(2,1)}
\label{sec:Legendre}

Legendre functions appear naturally in the representations of the rotation group SO(3) in three dimensions, the noncompact group SO(2,1), or of their covering group SU(2,$C$). See, for example, \cite{vilenkin,gilmore,biedenharn}. The Lie algebras so(3)$\,\simeq\,$su(2) are defined by three elements $J_1,\,J_2,\,J_3$ with Lie products given by the commutator $[J_1,J_2\,]=J_3$ and its cyclic permutations.  Thus, in a realization in which the Casimir operator $J_1^2+J_2^2+J_3^2$ has the fixed value $-\nu(\nu+1)$ and the commuting operator $J_3$ has the value $-i\mu$, the action of the so(3) algebra can be described in terms of coordinates $x_1=\sin\theta\cos\phi,\, x_2=\sin\theta\sin\phi,\ x_3=\cos\theta$ on the unit sphere $S^2$ by the action of the antiHermitian operators $J_1=-x_2\partial_3+x_3\partial_1$,
$J_2=-x_3\partial_1+x_1\partial_3$, $J_3=-x_1\partial_2+x_2\partial_1$ on the functions $e^{i\mu\phi}{\rm P}_\nu^\mu(\cos\theta)$. Here ${\rm P}_\nu^\mu(\cos{\theta})$ is the associated Legendre function ``on the cut'' $-1<\cos{\theta}<1$. This is defined in terms of the associated Legrendre function $P_\nu^\mu(z)$ for general complex $z$ by
\begin{equation}
\label{P_on_the_cut}
{\rm P}_\nu^\mu(\cos{\theta}) = \frac{1}{2}\left[e^{i\pi\mu/2} P_\nu^\mu(\cos{\theta}+i0) + e^{-i\pi\mu/2} P_\nu^\mu(\cos{\theta}-i0)\right],
\end{equation}
where $P_\nu^\mu(z)$ is given in terms of the hypergeometric function $_2F_1$  by  \cite{HTF}, Chap.\ 3,
\begin{equation}
\label{P_lm_def}
P_\nu^\mu(x) = \frac{1}{\Gamma(1-\mu)}\left(\frac{z+1}{z-1}\right)^{\mu/2}\,  _2F_1\left(-\nu,\nu+1;1-\nu\,;\frac{1-z}{2}\right)
\end{equation}
and its analytic continuations. The values of $\nu$ and $\mu$ are arbitrary. There is no restriction to the integer values characteristic of the associated Legendre polynomials unless one insists on a unitary representation of the group SO(3). We will not, and will simply be concerned with realizations of the algebra so(3). 

The algebra can also be realized on the functions $e^{i\mu\phi}{\rm Q}_\nu^\mu(\cos{\theta})$, with ${\rm Q}_\nu^\mu(\cos{\theta})$ a Legendre function of the second kind ``on the cut'',
\begin{equation}
\label{Q_on_the_cut}
{\rm Q}_\nu^\mu(\cos{\theta})= \frac{1}{2}e^{-i\pi\mu}\left[e^{-i\pi\mu/2} Q_\nu^\mu(\cos{\theta}+i0) + e^{-i\pi\mu/2}Q_\nu^\mu(\cos{\theta}-i0)\right],
\end{equation}
with $Q_\nu^\mu(z)$ defined for general complex $z$ by  
\begin{eqnarray}
\label{Q_lm_def}
Q_\nu^\mu(z) &=&  e^{i\pi\mu}2^{-\nu-1} \frac{\Gamma(\frac{1}{2})\Gamma(\nu+\mu+1)}{\Gamma(\nu+\frac{3}{2})} z^{-\nu-\mu-1}(z^2-1)^{\mu/2}
\nonumber
\\ 
&& \times\, _2F_1\left(1+\frac{\nu}{2}+\frac{\mu}{2},\frac{1}{2}+\frac{\nu}{2} +\frac{\mu}{2};\nu+\frac{3}{2};\frac{1}{z^2}\right).
\end{eqnarray}

The general forms of the $P$'s and $Q$'s appear naturally in representations of the noncompact group SO(2,1) on the unit hyperboloid $H^2$ with $x_1=\sinh\theta\cos\phi,\, x_2=\sinh\theta\sin\phi,\ x_3=\cosh\theta$, through the functions $e^{i\mu\phi}P_\nu^\mu(\cosh{\theta})$ and $e^{i\mu\phi}Q_\nu^\mu(\cosh{\theta})$, \cite{vilenkin}, Chap.~VI. SO(3) and SO(2,1) are different real forms of the covering group SO(3,$C$), and the Lie algebras are related \cite{gilmore}. It will be most convenient for our purposes to work with the general forms of the functions, and with realizations of so(2,1).  

The so(2,1) algebra is given in terms of three operators which we will take in the form
\begin{eqnarray}
\label{so(2,1)algebra}
M_1 &=& x_3\partial_1+x_1\partial_3, \nonumber \\
M_2 &=& x_3\partial_2+x_2\partial_3 \\
M_3 &=& x_2\partial_1-x_1\partial_2. \nonumber
\end{eqnarray}
These have the commutation relations
\begin{equation}
\label{so(2,1)commutators}
[M_1,M_2]=-M_3,\qquad [M_2,M_3]=M_1 ,\qquad [M_3,M_1]=M_2.
\end{equation}
$M_1$ and $M_1$ generate Lorentz transformations in the 1 and 2 directions, equivalent to hyperbolic rotations on $H^2$, while $M_3$ generates rotations in the 1,2 plane. 

The operator $-M_1^2-M_2^2+M_3^2$ is a Casimir invariant and may be taken to have fixed value on realizations of the algebra. We will also fix the value of the second commuting operator $iM_3$. When written in terms of the variables $z=\cosh{\theta}$ and $t=e^{i\phi}$, the relations $(-M_1^2-M_2^2+M_3^2)f = -\nu(\nu+1)f$, $iM_3f=\mu f$ imply that $f=f_\nu^\mu(z,t)=t^\mu F_\nu^\mu(z)$ where $F_\nu^\mu$ is a solution of the associated Legendre equation
\begin{equation}
\label{Legendre_eq}
\left[(1-z^2)\frac{d^2}{dz^2}-2z\frac{d}{dz}-\frac{\mu^2}{1-z^2} +\nu(\nu+1) \right]F_\nu^\mu(z) = 0
\end{equation}
with degree $\nu$ and order $\mu$

The so(2,1) algebra can be put in a more useful form for our purposes by introducing operators $M_\pm$ defined by
\begin{equation}
\label{M_pm}
M_\pm = \mp M_1 - iM_2
\end{equation}
with the commutation relations
\begin{equation}
\label{Mpm_commutators}
[iM_3,M_\pm]=\pm M_\pm,\qquad [M_+,M_-]=-2iM_3.
\end{equation}
In terms of the coordinates on $H^2$,
\begin{eqnarray}
\label{Mpm_coord}
M_+ &=& -e^{i\phi}\left(\partial_\theta + i\coth{\theta}\,\partial_\phi\right) =-t\sqrt{z^2-1}\,\partial_z+\frac{z}{\sqrt{z^2-1}}t^2\,\partial_t, 
\\
M_-&=& e^{-i\phi}\left(\partial_\theta-i\coth{\theta}\,\partial_\phi\right) = \frac{1}{t}\sqrt{z^2-1}\,\partial_z+\frac{z}{\sqrt{z^2-1}}\,\partial_t.
\end{eqnarray}

The commutation relations of $M_\pm$ with $M_3$ imply that $M_\pm t^\mu P_\nu^\mu(z) \!\propto \!t^{\mu\pm 1} P_\nu^{\mu\pm 1}(z)$ and $M_\pm t^\mu Q_\nu^\mu(z) $ $\propto t^{\mu\pm 1} Q_\nu^{\mu\pm 1}(z)$. The constants of proportionality are easily determined and are the same for $P_\nu^\mu$ and $Q_\nu^\mu$. After the $t$ dependence is extracted, these relations reduce to the standard differential recurrence relations for the order $\mu$, 
\begin{eqnarray}
\label{Pnumu_recurrence1}
-\sqrt{z^2-1}\frac{d}{dz}F_\nu^\mu(z)+\frac{\mu z}{\sqrt{z^2-1}} F_\nu^\mu(z)&=&-F_\nu^{\mu+1}(z), 
\\
\label{Pnumu_recurrence2}
\sqrt{z^2-1}\frac{d}{dz}F_\nu^\mu(z)+\frac{\mu z}{\sqrt{z^2-1}} F_\nu^\mu(z)&=&(\nu+\mu)(\nu-\mu+1)F_\nu^{\mu-1}(z),
\end{eqnarray}
where $F_\nu^\mu$ is a general solution of the associated Legendre equation,  \cite{HTF}, \S 3.8.


\section{Conformal extension of SO(2,1)}
\label{sec:conformal_extension}

We have so far dealt with SO(3) and SO(2,1) considered as the groups of transformations on $S^2$ and $H^2$. These appear as subgroups of the group of Euclidean transformations E(3), and of the group of Poincar\'{e} or pseudo-Euclidean transformations E(2,1), and are obtained by adding the translation operators in 3 or 2+1 dimensions to the original algebras. We will deal only with E(2,1). This is defined by the operators $M_i$ in \ref{so(2,1)algebra} and three translation operators $P_i=\partial_i$. We choose the metric such that $P^2=-P_1^2-P_2^2+P_3^2$, $M^2=-M_1^2-M_2^2+M_3^2$, and $x^2=-x_1^2-x_2^2+x_3^3$. 

The $P$'s commute,
\begin{equation}
\label{PiPjcomm}
[P_i,P_j]=0, 
\end{equation}
and transform as Lorentz vectors, with the commutation relations
\begin{equation}
\label{e(2,1)relations1} 
\begin{array}{lll}
\ [M_1,P_1]=-P_3, & \qquad [M_1,P_2]=0, & \qquad [M_1,P_3]=-P_1 \\ 
\ [M_2,P_1 ]=0 & \qquad [M_2,P_2]=-P_3, & \qquad [M_2,P_3]=-P_2 \\
\ [M_3,P_1]=P_2 & \qquad [M_3,P_2]=-P_1, & \qquad [M_3,P_3]=0.
\end{array}
\end{equation}
with repect to the generators $M_1$, $M_2$ of Lorentz transformations, and the generator $M_3$ of rotations. $P^2$ commutes with the $M$'s, and the solutions of the Klein-Gordon equation $P^2f=m^2f$ can be classified with respect to SO(2,1) by the values of $M^2$ and $M_3$.

In the special case that $P^2=0$, the symmetry group can be enlarged by the addition of a set of special conformal transformations with generators $K_i$ and the dilatation operator $D$. See, for example, \cite{miller2}, Chap.~4. These are given in terms of the coordinates $x_i$ by
\begin{eqnarray}
K_1 &=& 2x_1\,x\cdot\partial + x^2\partial_1 + x_1, \nonumber
\\
\label{Kdefinitions}
K_2 &=& 2x_2\,x\cdot\partial + x^2\partial_2 + x_2, 
\\
K_3 &=& -2x_3\,x\cdot\partial + x^2\partial_3 - x_3, \nonumber
\\
D &=& x\cdot\partial+\frac{1}{2}=x_1\partial_1+x_2\partial_2+x_3\partial_3 + \frac{1}{2}.
\end{eqnarray}
The $K$'s commute,
\begin{equation}
\label{Ks_commute}
[K_i,K_j]=0,
\end{equation}
and transform as Lorentz vectors,
\begin{equation}
\label{Kcommutators}
\begin{array}{lll}
\ [M_1,K_1]=-K_3, & \qquad [M_1,K_2]=0, & \qquad [M_1,K_3]=-K_1 \\
\ [M_2,K_1 ]=0 & \qquad [M_2,K_2]=-K_3, & \qquad [M_2,K_3]=-K_2 \\
\ [M_3,K_1]=K_2 & \qquad [M_3,K_2]=-K_1, & \qquad [M_3,K_3]=0.
\end{array}
\end{equation}
In addition, 
\begin{eqnarray}
\ [P_1,K_1]=[P_2,K_2]=2D, && \qquad  [P_3,K_3]=-2D \nonumber
\\
\label{[P,K]}
\ [P_1,K_2]=[K_1,P_2]=2M_3, && \qquad [P_3,K_i]=[K_3,P_i]=2M_i,\ \ i=1,2.
\end{eqnarray}
Finally, 
\begin{equation}
\label{[D,P][D,K]}
\ [D,P_i]=-P_i,  \qquad [D,K_i]=K_i, \qquad [D,M_i]=0,\ \ i=1,2,3.
\end{equation}

Using the explicit realization of the operators given above, we find also that 
\begin{eqnarray}
\label{P2=0_relations}
\ [K_i,P^2] &=& -4x_1P^2 \simeq 0, 
\\
\ [D,P^2] &=& 2P^2 \simeq 0,
\\
\label{MDcondition}
\ M^2+D^2 -\frac{1}{4} &=& x^2P^2 \simeq 0, 
\end{eqnarray}
where the final weak equivalence in each relation holds for the action of the operator on solutions $h$ of the wave equation $P^2h=0$. 

We will deal with the solutions of the wave equation in terms of the homogeneous functions
\begin{equation}
\label{hnumu_def}
h_\nu^\mu=x^\nu f_\nu^\mu(z,t)=x^\nu t^\mu F_\nu^\mu(z),
\end{equation}
where $t^\mu F_\nu^\mu(z)$ is a solution of the associated Legendre equation on $H^2$ with $M^2=-\nu(\nu+1)$ and $iM_3=\mu$ as in \ref{Legendre_eq}, and $x$, $t$, and $z$ are defined as 
\begin{equation}
\label{zt_def}
x=\sqrt{x^2},\qquad z=x_3/x, \qquad t=(x_1+ix_2)/\sqrt{x_1^2+x_2^2}. 
\end{equation}
In accord with the last two of equations \ref{P2=0_relations}, $Dh_\nu^\mu=(\nu+\frac{1}{2})h_\nu^\mu$, a relation which provides a geometric interpretation of the degree $\nu$ of the Legendre function. 

The operators $P_3$ and $K_3$ act as stepping operators in $\nu$. Thus, from the first two relations in \ref{[D,P][D,K]},
\begin{eqnarray}
\label{nu_stepping}
D(P_3 h_\nu^\mu) &=& P_3(D-1)h_\nu^\mu = (\nu-\frac{1}{2})P_3h_\nu^\mu, 
\\
D(K_3 h_\nu^\mu) &=& K_3(D+1)h_\nu^\mu = (\nu+\frac{3}{2})K_3h_\nu^\mu,
\end{eqnarray}
so $P_3h_\nu^\mu\propto h_{\nu-1}^\mu$ and $K_3h_\nu^\mu\propto h_{\nu+1}^\mu$.
In terms of our coordinates on $H^2$,
\begin{equation}
\label{P3coord}
P_3 = -\sinh^2{\theta}\frac{1}{x} \frac{\partial}{\partial\cosh{\theta}} + \cosh{\theta}\frac{\partial}{\partial x} 
= -(z^2-1) \frac{1}{x}\frac{\partial}{\partial z}+z\frac{\partial}{\partial x},
\end{equation}
while
\begin{eqnarray}
K_3 &=&  -x\sinh^2{\theta}\,\frac{\partial}{\partial\cosh{\theta}} -x^2\cosh{\theta}\left(\frac{\partial}{\partial x}+\frac{1}{x}\right)\nonumber
\\
\label{K3coord}
 &=&  -x(z^2-1)\frac{\partial}{\partial z}-x^2 z\left(\frac{\partial}{\partial x}+\frac{1}{x}\right).
\end{eqnarray}

Upon applying $P_3$ and $K_3$ to the functions $h_\nu^\mu$ defined in \ref{hnumu_def}, we find that
\begin{eqnarray}
P_3h_\nu^\mu(z) &=& x^{\nu-1}t^\mu \left[-(z^2-1)\frac{\partial}{\partial z} + \nu z\right]F_\nu^\mu(z) \nonumber
\\
\label{P3step}
 &=& x^{\nu-1}t^\mu (\nu+\mu)F_{\nu-1}^\mu(z), 
\\
K_3h_\nu^\mu &=& x^{\nu+1}t^\mu\left[-(z^2-1)\frac{\partial}{\partial z}-(\nu+1)z\right]F_\nu^\mu(z)
\nonumber
\\
\label{K3step}
&=&-x^{\nu+1}t^\mu(\nu-\mu+1)F_{\nu+1}^\mu(z).
\end{eqnarray}
These relations give the standard differential recurrence relations in $\nu$ for the associated Legendre functions $F_\nu^\mu$ after the dependence on $x$ and $t$ is removed. The indicated coefficients can be determined using the asymptotic behavior of those functions.


\section{Action of the group operators and generating functions}
\label{sec:SO3_action}

The action of the finite group operators $e^{-uM_\pm}$ on the functions $f^\mu_\nu$ is easily determined. Thus, writing the stepping operation by $M_+$ in the form
\begin{eqnarray}
\label{M+action}
M_+f_\nu^\mu &=& \left[-t\sqrt{z^2-1}\,\partial_z + \frac{z}{\sqrt{z^2-1}} t^2\partial_t\right]\,t^\mu F^\mu_\nu(z) \nonumber 
\\
&=& -t^{\mu+1}(z^2-1)^{(\mu+1)/2}\frac{d}{dz}\left[(z^2-1)^{-\mu/2}F_\nu^\mu(z) \right],
\end{eqnarray}
and noting that 
\begin{equation}
M_+t^\mu(z^2-1)^{\mu/2}=0,
\end{equation}
we can easily show that\footnote{This result is also easily obtained starting with the expression $t^\mu F_\nu^\mu(z)=(x_1+ix_2)^\mu \times(x_3^2-x^2)^{-\mu/2} F_\nu^\mu(x_3/x)$, writing $M_+$ as $-x_3(\partial_1+i\partial_2) -(x_1+ix_2)\partial_3$, and determining the action of $e^{-uM_+}$ directly using the relations $M_+(x_1+ix_2)=0$ and  $M_+x^2=0$. See also \cite{vilenkin}, Chap.~VI.} 
\begin{eqnarray}
\label{M+action2}
e^{-uM_+}t^\mu F_\nu^\mu(z) &=& t^\mu (z^2-1)^{\mu/2} e^{ut\sqrt{z^2-1}(d/dr)} (r^2-1)^{-\mu/2}F_\nu^\mu(r)\left.\right|_{r=z}
\nonumber 
\\
&=& t^\mu (z^2-1)^{\mu/2}\left[(z+ut\sqrt{z^2-1})^2-1\right]^{-\mu/2} F_\nu^\mu(z+ut\sqrt{z^2-1}).
\end{eqnarray}
Alternatively, by direct expansion of the exponential and the use of \ref{Pnumu_recurrence1},
\begin{equation}
\label{M+action3} 
e^{-uM_+}t^\mu F_\nu^\mu(z) = \sum_{n=0}^\infty \frac{u^n}{n!}t^{\mu+n} F_\nu^{\mu+n}(z).
\end{equation}
Comparison of the two expressions gives the generating function
\begin{eqnarray}
\label{M+generating}
t^\mu (z^2-1)^{\mu/2} &&\hspace*{-1em} \left[(z+ut\sqrt{z^2-1})^2-1\right]^{-\mu/2} F_\nu^\mu(z+ut\sqrt{z^2-1}) \nonumber
\\
&=& \sum_{n=0}^\infty \frac{u^n}{n!}t^{\mu+n} F_\nu^{\mu+n}(z).
\end{eqnarray}
The series converges for $|u|$ sufficiently small as expected from Lie theory, with absolute convergence for either $P_\nu^\mu$ or $Q_\nu^\mu$ for
\begin{equation}
\label{u_limits}
|u|<\frac{1}{|t|}\,{\rm min}\left|\sqrt{\frac{z\pm 1}{z\mp 1}}\right|.
\end{equation}
The generating function for $F_\nu^\mu=P_\nu^\mu$ is known, \cite{HTF},   19.10(4).

A similar calculation using $M_-$ gives the relations
\begin{eqnarray}
\label{M-action}
e^{-uM_-}t^\mu F_\nu^\mu(z) &=& \sum_{n=0}^\infty \frac{(-u)^n}{n!}t^{\mu-n}\frac{\Gamma(\mu+\nu+1) \Gamma(\nu-\mu+n+1)}{\Gamma(\mu+\nu-n+1) \Gamma(\nu-\mu+1)} F_\nu^{\mu-n}(z) \nonumber
\\
\label{M-action2}
&=& 
t^\mu (z^2-1)^{-\mu/2}\left[(z-\frac{u}{t}\sqrt{z^2-1})^2-1\right]^{\mu/2} F_\nu^\mu\left(z-\frac{u}{t}\sqrt{z^2-1}\right),
\end{eqnarray}
with absolute convergence of the series for the the condition in \ref{u_limits} with $1/t\rightarrow t$. 

The action of the finite operators $e^{-uP_3}$ and $e^{-uK_3}$ in the conformal group on the functions $h_\nu^\mu$ is well known \cite{miller2}, but it is useful to determine it directly. In order to use the method sketched above for $M_\pm$, we change from the variable $z=x_3/x$ to a new variable $y=x_3/(x_1^2+x_2^2)^{1/2} =z/\sqrt{z^2-1} = \coth{\theta}$ in terms of which
\begin{eqnarray}
\label{P3_y}
P_3 &=& \sqrt{y^2-1}\left(\frac{1}{x}\frac{\partial}{\partial y} + \frac{y}{y^2-1}\frac{\partial}{\partial x}\right), 
\\
\label{K3_y}
K_3 &=& \sqrt{y^2-1}\left(x\frac{\partial}{\partial y}-\frac{y}{y^2-1} x\frac{\partial}{\partial x}x\right).
\end{eqnarray}
In this form,
\begin{equation}
\label{PKfactors}
P_3\,x^\nu(y^2-1)^{\nu/2}=0, \qquad K_3\,x^\nu(y^2-1)^{(\nu+1)/2}=0,
\end{equation}
for any $\nu$. Extracting the relevant factors from the respective operands, we find that the actions of $P_3$ and $K_3$ can be written in terms of simple derivatives,
\begin{eqnarray}
P_3\,x^\nu F_\nu^\mu \left(\frac{y}{\sqrt{y^2-1}}\right) &=& P_3\,x^\nu(y^2-1)^{-\nu/2}\left[ (y^2-1)^{\nu/2}F_\nu^\mu\left(\frac{y}{\sqrt{y^2-1}}\right)\right] \nonumber
\\
\label{P3F}
&=& x^{\nu-1}(y^2-1)^{-(\nu-1)/2}\frac{d}{dy}\left[ (y^2-1)^{\nu/2}F_\nu^\mu \left(\frac{y}{\sqrt{y^2-1}}\right)\right],
\\
K_3\,x^\nu F_\nu^\mu \left(\frac{y}{\sqrt{y^2-1}}\right) &=& K_3\,x^\nu(y^2-1)^{(\nu+1)/2}\left[ (y^2-1)^{-(\nu+1)/2}F_\nu^\mu \left(\frac{y}{\sqrt{y^2-1}}\right)\right] \nonumber
\\
\label{K3F}
&=& x^{\nu+1}(y^2-1)^{(\nu+2)/2}\frac{d}{dy}\left[ (y^2-1)^{-(\nu+1)/2}F_\nu^\mu \left(\frac{y}{\sqrt{y^2-1}}\right)\right].
\end{eqnarray}
The use of $y$ or $\coth{\theta}$ as the the preferred variable is in this sense natural.

The relations above connect the action of $e^{-uP_3}$ and $e^{-uK_3}$ to Taylor series in $y$, and lead to the expressions 
\begin{eqnarray}
\label{eP3action}
\quad e^{-uP_3}x^\nu F_\nu^\mu &=& x^\nu\left(\frac{Y^2-1}{y^2-1}\right)^{\nu/2}F_\nu^\mu\left(\frac{Y}{\sqrt{Y^2-1}}\right), \quad Y=y-\frac{u}{x}\sqrt{y^2-1},
\\
\label{eK3action}
\quad e^{-uK_3}x^\nu F_\nu^\mu &=& x^\nu \left(\frac{y^2-1}{Y^2-1}\right)^{(\nu+1)/2}F_\nu^\mu\left(\frac{Y}{\sqrt{Y^2-1}}\right), \quad Y=y-ux\sqrt{y^2-1}. 
\end{eqnarray}
Equivalently, in terms of $z=\cosh{\theta}$ and the formal power series expansions of the exponentials,
\begin{eqnarray}
\quad e^{-uP_3}x^\nu F_\nu^\mu(z) &=& x^\nu \sum_{n=0}^\infty \frac{1}{n!}\left(-\frac{u}{x}\right)^n\frac{\Gamma(\nu+\mu+1)}{\Gamma(\nu +\mu-n+1)}F_{\nu-n}^\mu(z) \nonumber
\\ 
\label{eP3action2}
&=& x^\nu\left(1-2z\frac{u}{x}+\frac{u^2}{x^2}\right)^{\nu/2}F_\nu^\mu\left( \frac{z- \frac{u}{x}}{\sqrt{1-2z\frac{u}{x}+\frac{u^2}{x^2}}}\right),
\\
\quad e^{-uK_3}x^\nu F_\nu^\mu(z) &=& x^\nu \sum_{n=0}^\infty \frac{(ux)^n}{n!}\frac{\Gamma(\nu-\mu+n+1)}{\Gamma(\nu-\mu+1)}F_{\nu+n}^\mu(z) \nonumber
\\
\label{eK3action2}
&=& x^\nu\left(1-2zux+u^2x^2\right)^{-(\nu+1)/2}F_\nu^\mu\left( \frac{z- ux}{\sqrt{1-2zux+u^2x^2}}\right).
\end{eqnarray}
The series converge for $|h|<{\rm min}\,|z\pm\sqrt{z^2-1}|$, where $h=u/x$ for \ref{eP3action2}, and $h=ux$ for \ref{eK3action2}.  The expressions \ref{eP3action2} and \ref{eK3action2} reproduce the known generating functions \cite{HTF} 19.10(3) and 19.10(2), respectively, for $F_\nu^\mu=P_\nu^\mu$, but hold also for the functions $Q_\nu^\mu$. 

These relations give generating functions for the Legendre functions in terms of the degree $\nu$. Thus, starting with $P_0^0(z)=1$ and using \ref{eK3action2}, we obtain the standard generating function for the Legendre polynomials,
\begin{equation}
\label{Pngenerating}
e^{-uK_3}\cdot 1 = (1-2zux+u^2x^2)^{-1/2} = \sum_{n=0}^\infty (ux)^n P_n(z).
\end{equation}
The known generating function \cite{HTF} 3.7(34) for the $Q_n$ follows from \ref{eK3action2} with $Q_0^0\equiv Q_0=\frac{1}{2}\ln{(z+1)/}$ ${(z-1)}$,
\begin{eqnarray}
e^{-uK_3}\frac{1}{2}\ln{\frac{z+1}{z-1}} &=& (1-2zux+u^2x^2)^{-1/2}\ln{\left(\frac{z-ux\sqrt{1-2zux+u^2x^2}}{\sqrt{z^2-1}} \right)} \nonumber
\\
\label{Qngenerating}
&=& \sum_{n=0}^\infty (ux)^n Q_n(z). 
\end{eqnarray}

Other simple generating functions can be obtained using different starting points. Thus, using
\begin{equation}
\label{P0mu}
P_0^\mu(z) = \frac{1}{\Gamma(1-\mu)}\left(\frac{z+1}{z-1} \right)^{\mu/2}
\end{equation}
in \ref{eK3action2} gives
\begin{eqnarray}
e^{-uK_3}P_0^\mu(z) &=& \frac{1}{\Gamma(1-\mu)}(1-2zux+u^2x^2)^{-1/2} \left(\frac{z-ux+\sqrt{1-2zux+u^2x^2}}{\sqrt{z^2-1}}\right)^\mu \nonumber
\\
\label{Pnmugenerating}
&=&
\sum_{n=0}^\infty \frac{(ux)^n}{n!}\frac{\Gamma(-\mu+n+1)}{\Gamma(-\mu+1)} P_n^\mu(z). 
\end{eqnarray}
A similar generating function for the functions $Q_n^\lambda(z)$ follows from \ref{eK3action2} and the relation
\begin{equation}
\label{Q0mu} 
Q_0^\mu(z) = \frac{1}{2}e^{i\pi\mu}\Gamma(\mu)\left[\left(\frac{z+1}{z-1}\right)^{\mu/2} - \left(\frac{z+1}{z-1}\right)^{-\mu/2}\right].
\end{equation}
 Further results can be obtained using the known closed-form expressions for $Q_\nu^{1/2}$, $P_\nu^{1/2}$, $P_\nu^{-\nu}(z)$, and $Q_\nu^{\nu+1}(z)$, \cite{HTF}, \S 3.61.

We can also obtain interesting double series. Thus,
\begin{eqnarray}
e^{-vM_+}e^{-uK_3}\cdot 1 &=& \left[1-2ux\left(z+vt\sqrt{z^2-1}\right) + u^2x^2 \right]^{-1/2} \nonumber
\\
\label{Pnmseries}
&=& \sum_{n=0}^\infty(ux)^n\sum_{m=0}^n\frac{(vt)^m}{m!}P_n^m(z).
\end{eqnarray}
A series for $Q_n^m$ with the same coefficients follows from \ref{Qngenerating} with the replacement of $z$ by $z+vt\sqrt{v^2-1}$ in the function to be expanded. In that case, the sum on $m$ does not terminate.


\section{Fractional stepping operators from so(2,1)}
\label{sec:so(2,1)frac_ops}

We can define fractional stepping operators $M_\pm^\lambda$ as in \S \ref{sec:frac_op}, with
\begin{equation}
\label{Mpm^lambda}
M_\pm^\lambda = \frac{1}{2\pi i}e^{i\pi\lambda}\Gamma(\lambda+1)\int_{C} du\, \frac{e^{-uM_\pm}}{u^{\lambda+1}},
\end{equation}
where $C$ is an appropriately chosen contour in the complex $u$ plane. The operator $(iM_3)^\sigma$ is defined similarly.

The formal properties of the operators are easily determined. $[M^2,M_\pm^\lambda]=0$, so $M_\pm^\lambda$ transform solutions $f_\nu^\mu(z,t)=t^\mu F_\nu^\mu(z)$ of the associated Legendre equation $[M^2+\nu(\nu+1)]f_\nu^\mu=0$ into solutions with the same degree $\nu$. Further, from the relations
\begin{equation}
\label{M3Mpmcomm}
[iM_3,M_\pm^n]=\pm nM_\pm^n,
\end{equation}
we find that
\begin{equation}
\label{input}
[iM_3,e^{-uM_\pm}]=\pm\sum_{n=0}^\infty \frac{(-u)^n}{n!}nM_\pm^n = \pm u\frac{d}{du}e^{-uM_\pm},
\end{equation}
hence, after a partial integration in \ref{Mpm^lambda}, that
\begin{equation}
\label{JMcomm}
[iM_3,M_\pm^\lambda]=\pm\lambda M_\pm^\lambda.
\end{equation}
The operators $M_\pm^\lambda$ therefore increase or decrease the order $\mu$ by $\lambda$ when applied to $f_\nu^\mu=t^\mu F_\nu^\mu(z)$, 
\begin{equation}
\label{M_lambda_stepping}
iM_3(M_\pm^\lambda f_\nu^\mu)=M_\pm^\lambda(iM_3\pm\lambda)f_\nu^\mu = (\mu\pm\lambda)(M_\pm^\lambda f_\nu^\mu).
\end{equation}
The new functions $M_\pm^\lambda f_\nu^\mu$ may involve a different combination of the fundamental solutions of the associated Legendre equation than appeared originally, with $(f')_\nu^{\mu\pm\lambda}= t^{\mu\pm\lambda} (F')_\nu^{\mu\pm\lambda}$. Furthermore, even when $F'=F$, the operations only give $f'$ up to a constants of proportionality because of the sign and numerical factors in the recurrence relations \ref{Pnumu_recurrence1} and \ref{Pnumu_recurrence2}. We will therefore write $M_\pm^\lambda f_\nu^\mu$ as 
\begin{equation}
\label{Nnumu_defined}
M_\pm^\lambda f_\nu^\mu=N_\pm(\nu,\mu,\lambda)(f')_\nu^{\mu\pm\lambda},
\end{equation}
where the functional form of $F'$ and the constant $N$ are to be determined. 

The more general relations
\begin{equation}
\label{[Jsigma,Mlambda}
(iM_3)^\sigma M_\pm^\lambda=M_\pm^\lambda(iM_3\pm\lambda)^\sigma,\qquad M_\pm^\lambda(iM_3)^\sigma=(iM_3\mp\lambda)^\sigma M_\pm^\lambda,
\end{equation}
can be derived using the Baker-Hausdorff expansion of $e^Ae^Be^{-A}$ as a series of $n$-fold commutators \cite{FracOpsBessel}. The operators $M_\pm$ are mixed with $M_3$ under commutation as in Eq. \ref{so(2,1)commutators}, with the result that there are apparently no simple expressions for commutators of the fractional operators $M_+^\lambda$ and $M_-^\eta$ for arbitrary values of $\lambda$ and $\eta$. However,
\begin{equation}
\label{M+M-powers}
\qquad [iM_3,M_+^\sigma M_-^\lambda] = (\sigma-\lambda)M_+^\sigma M_-^\lambda, \qquad
[iM_3,M_-^\lambda M_+^\sigma] = (\sigma-\lambda)M_-^\lambda M_+^\sigma,
\end{equation}
so $M_+^\sigma M_-^\lambda$ and $M_-^\lambda M_+^\sigma$ both carry solutions  $f_\nu^\mu$ to solutions of the type $(f')_\nu^{\mu+\sigma-\lambda}$.

We can define fractional operators $P_3^\lambda$ and $K_3^\lambda$ as above.  Since $[P^2,P_3^\lambda]=0$ and $[P^2,K_3^\lambda]\simeq 0$, these operators transform solutions $h_\nu^\mu(x,z,t)=x^\nu f_\nu^\mu(z,t)$ of $P^2h_\nu^\mu=0$ into solutions. Furthermore, $[M_3,P_3^\lambda] = [M_3,K_3^\lambda]=0$, so $P_3^\lambda$ and $K_3^\lambda$ do not change the value $\mu$ of $iM_3$. 

Calculations similar to those which lead to \ref{JMcomm} show that  
\begin{equation}
\label{DPlambda}
[D,P_3^\lambda]=-\lambda P_3^\lambda, \qquad [D,K_3^\lambda]=\lambda K_3^\lambda,
\end{equation}
so these operators decrease or increase the value of $\nu$ by $\lambda$. Given the numerical factors in the recurrence relations \ref{P3step} and \ref{K3step}, the transformed functions $h'$ are determined only up to constants of proportionality,
\begin{equation}
\label{N'defined}
P_3^\lambda \,h_\nu^\mu = N'_-(\nu,\mu,\lambda)\,h_{\nu-\lambda}^\mu, \qquad K_3^\lambda \,h_\nu^\mu = N'_+(\nu,\mu,\lambda)\,h_{\nu+\lambda}^\mu.
\end{equation}

Because
\begin{equation}
\label{P3Mpm}
[M_\pm,P_3]=\pm P_\pm, \qquad [M_\pm,K_3]=\pm K_\pm,
\end{equation}
the operators $M_\pm^\sigma$ do not commute with $P_3^\lambda$ and $K_3^\lambda$. However, it is easily shown that $M_\pm^\sigma P_3^\lambda$ and $P_3^\lambda M_\pm^\sigma$ both carry $h_\nu^\mu$ to solutions of the type $(h')_{\nu-\lambda}^{\mu\pm\sigma}$. Similarly, $M_\pm^\sigma K_3^\lambda$ and $K_3^\lambda M_\pm^\sigma$ both carry $h_\nu^\mu$ to solutions of the type $(h')_{\nu+\lambda}^{\mu\pm\sigma}$.


\section{Change of the order of $F_\nu^\mu$ using $M_\pm^\lambda$}
\label{sec:change_of_order}

\subsection{Weyl-type relations using $M_+^\lambda$}
\label{subsec:M+Weyl}

The action of the Weyl-type operator $M_+^\lambda$ on the associated Legendre functions gives
\begin{eqnarray}
M_+^\lambda t^\mu F_\nu^\mu(z) &=& N_+(\nu,\mu,\lambda)\,t^{\mu+\lambda}(F')_\nu^{\mu+\lambda}(z) \nonumber
\\
\label{Legendre-W}
&=& \frac{1}{2\pi i} e^{i\pi\lambda}\Gamma(\lambda+1)t^\mu(z^2-1)^{\mu/2} 
\\
&&\times \int_{C_W} \frac{du}{u^{\lambda+1}}\left[ \left(z+ut\sqrt{z^2-1}\right)^2-1\right]^{-\mu/2}
F_\nu^\mu\left(z+ut\sqrt{z^2-1}\right),\nonumber
\end{eqnarray}
where the direction of the contour $C_W$ for $|u|\rightarrow\infty$ must be chosen to assure convergence of the integral. 

It is easily shown that the integral converges and gives a solution of the associated Legendre equation for the choice $F_\nu^\mu=Q_\nu^\mu$ provided ${\rm Re}(\nu+\mu+\lambda+1)>0$.\footnote{It can be established that the integrals \ref{Q-W} and \ref{P-W} below give solutions of the associated Legendre equation with the indicated indices by a calculation similar to that which leads to the condition \ref{M+condition} for the Riemann version of $M_+^\lambda$. The only change in the final condition is the replacement of $v$ by $-v$ in \ref{M+condition}, and the restrictions given on $\nu$, $\mu$, and $\lambda$ follow.} The proportionality of the integral to $Q_\nu^{\mu+\lambda}$ and the value of the coefficient $N_+$ can be established using the asymptotic behavior of the two sides of \ref{Q-W} for $z\rightarrow\infty$, proportional in both cases to $z^{-\nu-1}$. We find that $N_+=e^{-i\pi\lambda}$, a result consistent with repeated application of the recurrence relation $M_+t^\mu F_\nu^\mu =-t^{\mu+1}F_\nu^{\mu+1}$. 

After scaling out the variable $t$, \ref{Legendre-W} can be rewritten for $F_\nu^\mu=Q_\nu^\mu$ as
\begin{equation}
\label{Q-W}
\qquad e^{-i\pi(\mu+\lambda)}Q_\nu^{\mu+\lambda}(z) = \frac{e^{i\pi\lambda}}{2\pi i}\Gamma(\lambda+1) \int_{(\infty,0+,\infty)} \frac{du}{u^{\lambda+1}} \left( \frac{z^2-1}{Z^2-1}\right)^{\mu/2}
 e^{-i\pi\mu}Q_\nu^\mu(Z),
\end{equation}
where $Z=z+u\sqrt{z^2-1}$ and ${\rm Re}(\nu+\mu+\lambda+1)>0$. That is,
\begin{equation}
\label{Qraised}
M_+^\lambda t^\mu e^{-i\pi\mu}Q_\nu^\mu(z)=t^{\mu+\lambda}e^{-i\pi(\mu+\lambda)}Q_\nu^{\mu+\lambda}(z).
\end{equation}

The expression in \ref{Q-W} can be put in the form of a generalized Weyl fractional integral by changing to $Z$ as the integration variable,
\begin{eqnarray}
\label{Q-Wfrac} 
e^{-i\pi(\mu+\lambda)}Q_\nu^{\mu+\lambda}(z) &=& \frac{1}{2\pi i}e^{i\pi\lambda}\Gamma(\lambda+1) (z^2-1)^{(\nu+\lambda)/2} \nonumber 
\\
&& \times \int_{(\infty,z+,\infty)} \frac{dZ}{(Z-z)^{\lambda+1}}\left(Z^2-1\right)^{-\mu/2}e^{-i\pi\mu}Q_\nu^\mu(Z).
\end{eqnarray}
The contour can be collapsed for ${\rm Re}\,\lambda < 0$, and \ref{Q-W} can then reduces to a form equivalent to the known fractional integral \cite{TIT}, 13.2(30). However, the result in \ref{Q-W} is more general and has a clear connection with the underlying group theory.

The further substitutions $z=\cosh{\theta}$, $Z=\cosh{\theta'}$ convert \ref{Q-Wfrac} to an expression in terms of hyperbolic angles,
\begin{eqnarray}
e^{-i\pi(\mu+\lambda)}Q_\nu^{\mu+\lambda}(\cosh{\theta}) &=& \frac{1}{2\pi i} e^{i\pi\lambda}\Gamma(\lambda+1)\sinh^{\mu+\lambda}{\theta} \int_{(\infty,\theta +,\infty)}\frac{d\theta'}{(\cosh{\theta'}-\cosh{\theta})^{\lambda+1}}
\nonumber
\\
\label{Q-W_hyper}
&& \times \sinh^{-\mu+1}{\theta'} e^{-i\pi\mu} Q_\nu^\mu(\theta').
\end{eqnarray}

A different result holds for the Weyl action of $M_+$ on the functions $t^\mu P_\nu^\mu(z)$ as may be seen from the relation
\begin{equation}
\label{P=Q-Q}
P_\nu^\mu(z) = \frac{1}{\pi}e^{-i\pi\mu}\frac{1}{\cos{\pi\nu}}\left[ \sin{\pi(\nu+\mu)}Q_\nu^\mu(z)-\sin{\pi(\nu-\mu)}Q_{-\nu-1}^\mu(z)\right].
\end{equation}
The operator $M_+^\lambda$ acts on the $Q$'s, but does not change the sine factors, with the result that, suppressing the the dependence on $t$,
\begin{eqnarray}
\label{P-W}
M_+^\lambda P_\nu^\mu(z) &=& \frac{1}{2\pi i}e^{i\pi\lambda}\Gamma(\lambda+1) \int_{(\infty,0+,\infty)} \frac{du}{u^{\lambda+1}} \left(\frac{z^2-1}{Z^2-1}\right)^{\mu/2} P_\nu^\mu(Z) \nonumber \\
\label{P-W1}
&=& \frac{1}{\pi}e^{-i\pi(\mu+\lambda)}\frac{1}{\cos{\pi\nu}}\left[ \sin{\pi(\nu+\mu)}Q_\nu^{\mu+\lambda}(z)-\sin{\pi(\nu-\mu)} Q_{-\nu-1}^{\mu+\lambda}(z)\right]  \\
\label{P-W2}
&=&\frac{\sin{\pi(\nu-\mu)}}{\sin{\pi(\nu-\mu-\lambda)}} P_\nu^{\mu+\lambda}(z)-\frac{2}{\pi}\sin{\pi\nu}\sin{\pi\lambda}\, e^{-i\pi(\mu+\lambda)} Q_\nu^{\mu+\lambda}(z) \nonumber
\end{eqnarray}
for ${\rm Re}(\pm(\nu+\frac{1}{2})+\mu+\lambda+\frac{1}{2})>0$ and $Z=z+\sqrt{z^2-1}$.

We emphasize that the different behavior of the Weyl-type operator $M_+^\lambda$ on $Q_\nu^\mu$ and $P_\nu^\mu$ is associated with the fact that the integration contour runs to $\infty$. $Q_\nu^\mu$ and $Q_{-\nu-1}^\mu$ have unique asymptotic limits for $z\rightarrow\infty$, behaving respectively as $z^{-\nu-1}$ and $z^\nu$ multiplied by series in $1/z^2$. The operator $M_+^\lambda$ changes $\mu$ but does not affect $\nu$. Since it carries solutions of the associated Legendre equation to solutions and, as is evident from \ref{Legendre-W}, preserves the asymptotic behavior of the integrand for $z\rightarrow\infty$, it can only carry $Q_\nu^\mu$ to a multiple of $Q_\nu^{\mu+\lambda}$ and $Q_{\nu-1}^\mu$ to a multiple of $Q_{-\nu-1}^{\mu+\lambda}$ with no mixing of the two functions. $P_\nu^\mu$, in contrast, involves both functions with different $\mu$-dependent coefficients, and cannot be reproduced for general $\nu$. It is useful in this respect to regard $Q_\nu^\mu$ and $Q_{-\nu-1}^\mu$ as the fundamental solutions of the associated Legendre equation rather than $P_\nu^\mu$ and $Q_\nu^\mu$. We will encounter similar situations later.


\subsection{Weyl-type relations using $M_-^\lambda$}
\label{subsec:M-Weyl}

The action of the Weyl-type operators $M_-^\lambda$ on the associated Legendre functions gives
\begin{eqnarray}
M_-^\lambda t^\mu F_\nu^\mu(z) &=& N_-(\nu,\mu,\lambda) t^{\mu-\lambda} (F')_\nu^{\mu-\lambda}(z) \nonumber
\\
\label{Legendre-W2}
&=& \frac{1}{2\pi i}\Gamma(\lambda+1) e^{i\pi\lambda} t^\mu(z^2-1)^{-\mu/2} 
\\
&& \times \int_{C_W} \frac{du}{u^{\lambda+1}} \left[ \left(z-\frac{u}{t}\sqrt{z^2-1}\right)^2-1\right]^{\mu/2}
 F_\nu^\mu\left(z-\frac{u}{t}\sqrt{z^2-1}\right).\nonumber
\end{eqnarray}
It can be established through a calculation equivalent to that which leads to the condition \ref{M_condition} obtained later for the Riemann version of $M_-^\lambda$ that this expression gives a solution of the associated Legendre equation provided the integral converges for $|u|\rightarrow\infty$. 

The contour $C_W$ must extend to $|u|\rightarrow\infty$, and must avoid the singularities of the integrand at $u/t=\sqrt{(z-1)/(z+1)}$, $\sqrt{(z+1)/(z-1)}$. The singularities are always in the right half of the $u/t$ plane a finite distance from the origin for $|z\pm 1|$ finite. For definiteness, we will consider the case in which, after scaling out the variable $t$, the initial contour is taken to run above both singularities. This allows us to rotate the contour in \ref{Legendre-W2} counterclockwise by $\pi$, and then replace $u$ by $e^{i\pi}u$, effectively returning to the expression in \ref{Legendre-W2} with $-u$ replaced by $u$ and with the acquisition of an extra phase $e^{-i\pi\lambda}$.  

After identifying the constant of proportionality, we find for the choice $F_\nu^\mu=Q_\nu^\mu$ that
\begin{eqnarray}
e^{-i\pi(\mu-\lambda)}Q_\nu^{\mu-\lambda}(z) &=& 
\frac{1}{2\pi i}e^{i\pi\lambda}\Gamma(\lambda+1)\frac{\Gamma(\nu+\mu-\lambda+1) \Gamma(\nu-\mu+1)}{\Gamma(\mu+\nu+1) \Gamma(\nu-\mu+\lambda+1)}  \nonumber
\\
\label{Q-W3}
&& \times \int_{(\infty,0+,\infty)} \frac{du}{u^{\lambda+1}} \left(\frac{Z^2-1}{z^2-1}\right)^{\mu/2}
 e^{-i\pi\mu}Q_\nu^\mu(Z),
\end{eqnarray}
$Z=z+u\sqrt{z^2-1}$. The integral converges and gives a solution $Q_\nu^{\nu-\lambda}$ of the associated Legendre equation for ${\rm Re}(\nu-\mu+\lambda+1)>0$. 

The coefficient of the integral can be determined using the asymptotic behavior of the two sides of \ref{Q-W3} for $z\rightarrow\infty$. It corresponds to a coefficient
\begin{equation}
\label{N-}
N_-(\nu,\mu,\lambda)= e^{-i\pi\lambda}\frac{\Gamma(\nu+\mu+1)\Gamma(\nu-\mu+\lambda+1)} {\Gamma(\nu+\mu-\lambda+1)\Gamma(\nu-\mu+1)} 
\end{equation}
in the original expression \ref{Legendre-W2}, with the contour taken to run above the singularities of the integrand at $u/t=\sqrt{(z\pm 1)/(z\mp 1)}$. $N_-$ is just the coefficient of the $n$th term in the generating function \ref{M-action2}, up to the factor $1/n!$, but extended from integer $n$ to noninteger values $n\rightarrow\lambda$ with the factor $(-1)^n\rightarrow e^{-i\pi\lambda}$. We note that the phase of the constant $N_-$ defined through \ref{Legendre-W2} depends on the choice of the original contour. The final result does not.\footnote{\label{phase}The contour rotation and substitution used above replace the factor $\exp{(-uM_-)}$ in the definition of $M_-^\lambda$ by $\exp{(-u\,e^{i\pi}M_-)}$, effectively replacing $-u$ by $u$ in the series \ref{M-action2}. The resulting expression  gives $(e^{i\pi}M_-)^\lambda$ when the integration in \ref{Mpm^lambda} is performed on the standard Weyl contour $(\infty,0+,\infty)$, hence the extra phase noted above in the expression for $M_-^\lambda$. Had we started instead with a contour that ran below both the singularities of the integrand, rotated the contour clockwise by $\pi$, and then replaced $u$ by $e^{-i\pi}u$, the extra phase would have been $e^{i\pi\lambda}$, and the new $N_-$ would be $e^{2i\pi\lambda}$ times the result in \ref{N-}. The ultimate expression for $Q_\nu^{\mu-\lambda}$ does not change.} The contour in \ref{Q-W3} can be closed for $\lambda=n$, one recovers the expression for $Q_\nu^{\mu-n}$ given by the generating function \ref{M-action2}.

The results above continue to hold for $Q_{-\nu-1}^\mu$ for ${\rm Re}(-\nu-\mu+\lambda)>0$, and \ref{Q-W3} and \ref{P=Q-Q} can be used to evaluate $M_-^\lambda P_\nu^\mu$. Remarkably, the coefficients $N_-(\nu,\mu,\lambda)$ and $N_-(-\nu-1,\mu,\lambda)$ are such that the sine functions in \ref{P=Q-Q} are reproduced in the final result with $\mu\rightarrow\mu-\lambda$, and we find that
\begin{eqnarray}
P_\nu^{\mu-\lambda}(z) &=& 
\frac{1}{2\pi i}e^{i\pi\lambda}\Gamma(\lambda+1)\frac{\Gamma(-\nu-\mu) \Gamma(\nu-\mu+1)}{\Gamma(-\nu-\mu+\lambda) \Gamma(\nu-\mu+\lambda+1)}  \nonumber
\\
\label{P-W3}
&&  \times \int_{(\infty,0+,\infty)} \frac{du}{u^{\lambda+1}} \left( \frac{Z^2-1}{z^2-1}\right)^{\mu/2} P_\nu^\mu(Z),
\end{eqnarray}
${\rm Re}(\pm(\nu+\frac{1}{2})-\mu+\lambda+\frac{1}{2})>0$. This corresponds to a coefficient
\begin{equation}
\label{Ntilde_-}
\qquad \tilde{N}_-=e^{i\pi\lambda}\frac{\sin{\pi(\nu+\mu)}}{\sin{\pi(\nu+\mu-\lambda)}}N_-(\nu,\mu,\lambda)=\frac{\Gamma(-\nu-\mu+\lambda)\Gamma(\nu-\mu+\lambda+1)}{ \Gamma(-\nu-\mu)\Gamma(\nu-\mu+1)}
\end{equation}
in the relation $M_-^\lambda P_\nu^\mu=\tilde{N}_-(\nu,\mu,\lambda) P_\nu^{\mu-\lambda}$, with $M_-^\lambda$ defined with the original contour in \ref{Legendre-W2} taken above the singularities of the integrand. The change relative to \ref{N-} is in a different continuation of the alternating sign in the generating function \ref{M-action2}, with the factor $e^{-i\pi\lambda}$ in $N_-$ replaced by $\sin{\pi(\nu+\mu)}/\sin{\pi(\nu+\mu-\lambda)}$ in $\tilde{N}_-$. Because of the change in coefficients, it is again useful to regard $Q_\nu^\mu$ and $Q_{-\nu-1}^\mu$ as the fundamental solutions rather than $Q_\nu^\mu$ and $P_\nu^\mu$.


\subsection{Riemann-type relations using $M_\pm^\lambda$}
\label{subsec:RiemannLegendre}

The action of the Riemann-type fractional operator $M_+^\lambda$ is given by \ref{Legendre-W} on a finite contour $C_R=(u_0,0+,u_0)$ where the obvious choice of the endpoint $u_0$ is the point at which the argument of the Legendre function is $+1$,
\begin{equation}
\label{u0}
u_0 = e^{i\pi}\frac{1}{t}\left(\frac{z-1}{z+1}\right)^{1/2}.
\end{equation}
Writing $u$ as $u_0v$ and factoring out the dependence on $t$, we obtain the expression
\begin{eqnarray}
N_+\left(\frac{z-1}{z+1}\right)^{(\mu+\lambda)/2}\left(F'\right)_\nu^{\mu+\lambda}(z) &=& \frac{1}{2\pi i}\Gamma(\lambda+1) \int_{(1,0+,1)}\frac{dv}{v^{\lambda+1}} \frac{1}{(1-v)^\mu} \nonumber
\\
\label{Fmu+lambda_R}
&&\times  \left(\frac{V-1}{V+1}\right)^{\mu/2} F_\nu^\mu(V),
\end{eqnarray}
where $V=z-(z-1)v$ and $F$ and $F'$ are possibly different associated Legendre functions. The last factor on the right hand side has the same form as the function on the left, but with $z$ replaced by $V$. 

Defining
\begin{equation}
\label{w_munu}
w_\nu^\mu(z) = \left(\frac{z-1}{z+1}\right)^{\mu/2}F_\nu^\mu(z)
\end{equation}
and using the form of the associated Legendre equation satisfied by that function,
\begin{equation}
\label{diff_eq_w}
(z^2-1)w''+(2z-2\mu)w'-\nu(\nu+1)w=0,
\end{equation}
with the replacement $\mu\rightarrow\mu+\lambda$, we find that \ref{Fmu+lambda_R} gives a solution of \ref{diff_eq_w} provided
\begin{equation}
\label{M+condition}
v^{-\lambda}(1-v)^{1-\mu}\frac{d}{dv}\left[\left(\frac{V-1}{V+1}\right)^{\mu/2} F_\nu^\mu(V)\right]=0
\end{equation}
at the endpoints of the integration contour. This condition is satisfied for $P_\nu^\mu$ for ${\rm Re}\,\mu<1$ with the expected endpoints $v=1,\ e^{2\pi i}$. The expression in \ref{M+condition} does not vanish for $Q_\nu^\mu$, with the result that the right hand side of \ref{Fmu+lambda_R} satisfies an inhomogenous version of the associated Legendre equation, so does not give $Q_\nu^{\mu+\lambda}$.  

With the choice $F_\nu^\mu=P_\nu^\mu$ in \ref{Fmu+lambda_R}, we find that $N_+= e^{-i\pi\lambda}$ as before, and that
\begin{eqnarray}
\label{TIT13.1(54)-1}
P_\nu^{\mu+\lambda}(z) &=& \frac{1}{2\pi i} e^{2i\pi\lambda} \Gamma(\lambda+1)\int_{(u_0,0+,u_0)} \frac{du}{u^{\lambda+1}} 
\left(\frac{z^2-1}{Z^2-1}\right)^{\mu/2} P_\nu^\mu(Z), \nonumber
\\
\label{TIT13.1(54)}
&=& \frac{1}{2\pi i} e^{i\pi\lambda} \Gamma(\lambda+1)\int_{(u'_0,0+,u'_0)} \frac{du}{u^{\lambda+1}} 
\left(\frac{z^2-1}{Z^{'2}-1}\right)^{\mu/2} P_\nu^\mu(Z'),
\end{eqnarray}
where ${\rm Re}\,\mu<1$, $Z'=z-u\sqrt{z^2-1}$, and $u'_0=\sqrt{(z-1)/(z+1)}$. Changing to $Z'$ as the integration variable, we get the alternative form
\begin{eqnarray}
P_\nu^{\mu+\lambda}(z) &=& \frac{1}{2\pi i}  \Gamma(\lambda+1) (z^2-1)^{(\mu+\lambda)/2} \nonumber
\\
\label{TIT13.1(54)-2}
&& \times \int_{(1,z+,1)} \frac{dZ'}{(Z'-z)^{\lambda+1}} 
(Z^{'2}-1)^{-\mu/2} P_\nu^\mu(Z'),
\end{eqnarray}
where $|{\rm arg}(Z'-z)|\leq\pi$. This result can be reduced in the case of real $z$ with $-1<z<1$ to a known the fractional integral, \cite{TIT}, 13.1(54). 

A similar calculation for $M_-^\lambda$ on the Riemann contour starting from \ref{Legendre-W2} leads to the condition for a solution of the associated Legendre equation that the function
\begin{equation}
\label{M_condition}
v^{-\lambda}(1-v)^{\mu+1}\frac{d}{dv}\left[\left(\frac{V+1}{V-1}\right)^{\mu/2} F_\nu^\mu\left(V)\right)\right]
\end{equation}
vanish at the end points. In fact, it has a finite value for $v=1,\ e^{2\pi i}$ for either $P_\nu^\mu$ or $Q_\nu^\mu$. The functions defined by the integral satisfy inhomogeneous versions of the associated Legendre equation rather than the equation itself, and the Riemann version of $M_-^\lambda$ appears not to be useful.


\section{Change of the degree of $F_\nu^\mu$ using $K_3^\lambda$ and $P_3^\lambda$} 
\label{sec:K3,P3^lambda}

\subsection{Relations for $K_3^\lambda$}
\label{subsec:K3^lambda_Weyl}

The action of the operator $K_3^\lambda$ on a function $x^\nu F_\nu^\mu$ increases the degree $\nu$ by $\lambda$ as shown formally by the commutation relation \ref{DPlambda}. In particular, using the variable $y$ and the expression in \ref{eK3action} for the action of $e^{-uK_3}$, we obtain a Weyl-type relation
\begin{eqnarray}
K_3^\lambda x^\nu F_\nu^\mu\left(\frac{y}{\sqrt{y^2-1}}\right) &=& N'_+(\nu,\mu,\lambda)x^{\nu+\lambda}
\left(F'\right)_{\nu+\lambda}^\mu\left(\frac{y}{\sqrt{y^2-1}}\right) \nonumber \\
\label{K3lambdaF}
&& \hspace*{-30pt}=\frac{1}{2\pi i}e^{i\pi\lambda}\Gamma(\lambda+1) 
\int_{C_W}\frac{du}{u^{\lambda+1}}\left(\frac{y^2-1}{Y^2-1} \right)^{(\nu+1)/2} F_\nu^\mu\left(\frac{Y}{\sqrt{Y^2-1}}\right), 
\end{eqnarray}
where $Y=y-xu\sqrt{y^2-1}$. The contour $C_W$ must be chosen to run to $|u|\rightarrow\infty$, avoiding the singularities of the integrand at $Y=\pm 1$ or $xu=\sqrt{(y\pm 1)/(y\mp 1)}$. The singularities are both in the right-half $xu$ plane a finite distance from the origin for $|y\pm 1|$ finite. 

Following the procedure sketched in \S\ref{subsec:M-Weyl}, we change to $xu$ as a new integration variable, pick an initial contour which runs above both singularities, rotate the contour counterclockwise by $\pi$, and replace $u$ by $e^{i\pi}u$ to return to the standard contour. This gives the expression 
\begin{eqnarray}
N'_+(\nu,\mu,\lambda)\left(F'\right)_{\nu+\lambda}^\mu\left(\frac{y}{\sqrt{y^2-1}}\right) &=& \frac{1}{2\pi i}\Gamma(\lambda+1) \int_{(\infty,0+,\infty)}\frac{du}{u^{\lambda+1}} \nonumber
\\
\label{K3lambdaF-1}
&& \times  \left(\frac{y^2-1}{Y^2-1} \right)^{(\nu+1)/2} F_\nu^\mu\left(\frac{Y}{\sqrt{Y^2-1}}\right),
\end{eqnarray}
where $Y$ is now given by $Y=y+u\sqrt{y^2-1}$ and we have acquired an extra phase $e^{-i\pi\lambda}$. By considering the modified Legendre equation
\begin{equation}
\label{vnumu-eqn}
(y^2-1)v''+\left(2y-\nu-\frac{1}{2}\right)v'-\left(\mu^2-\frac{1}{4}\right)v=0
\end{equation}
satisfied by the function\footnote{The form of the function is suggested by the Whipple transformation \cite{HTF}3.3.2(13,14)
\begin{displaymath}
e^{-i\pi\mu}Q_\nu^\mu(\frac{y}{\sqrt{y^2-1}})=\sqrt{\frac{\pi}{2}}\Gamma(\nu+\mu+1)(y^2-1)^{1/4}P_{-\mu-\frac{1}{2}}^{-\nu-\frac{1}{2}}(y)
\end{displaymath}
which connects the actions of $K_3^\lambda$ and $M_-^\lambda$ and of $P_3^\lambda$ and $M_+^\lambda$, and can be used to derive equations \ref{K3lambdaF-2} and \ref{P3lambdaF-2} below from \ref{P-W3} and \ref{Q-W}. 
The Whipple transformation is associated with the automorphism
\begin{eqnarray}
D'=-iM_3,\quad iM_3'&=&-D,\quad M_+'=P_3,\quad M_-'=-K_3 ,\nonumber
\\
P_+'=-P_+,\quad P_-'=K_+,\quad  P_3'&=&M_+,\quad K_+'=P_-,\quad K_-'=-K_-,\quad K_3'=-M_- \nonumber
\end{eqnarray}
of the abstract conformal algebra. Using the explicit realizations of the operators given in \S\ref{sec:conformal_extension}, we find that
\begin{displaymath}
P^{'2}=x_+^2P^2\simeq 0 \quad {\rm and}\quad M^{'2}+D^{'2}-\frac{1}{4}=-\frac{1}{2}(P_3K_3+K_3P_3)-D^2-M_3^2-\frac{1}{4}= x_\perp^2P^2\simeq 0
\end{displaymath}
when acting on solutions of the Laplace equation. These relations are only invariant under rotations about the 3 axis generated by $M_3$ and not under the full Lorentz group. The transformation maps realizations of the algebra to realizations, with $\nu'=-\mu-\frac{1}{2}$ ($iM_3'=-D$) and $\mu'=-\nu-\frac{1}{2}$ ($D'=-iM_3$). In terms of differential equations, the substitutions $z = y/\sqrt{y^2-1}$, $F_\nu^\mu(z)=(y^2-1)^{1/4}G_{\nu'}^{\mu'}(y)$ transform the associated Legendre equation satisfied by $F_\nu^\mu(z)$ into the same equation for $G_{\nu'}^{\mu'}(y)$. The specific connection of the functions can be established by matching their behavior for $z\rightarrow\infty,\ y\rightarrow 1$ and for $z\rightarrow 1,\ y\rightarrow\infty$.
}
\begin{equation}
\label{wmunu}
v_\nu^\mu(y) = \left(\frac{y-1}{y+1}\right)^{(\nu+1)/2}(y^2-1)^{-1/4} F_\nu^\mu \left(\frac{y}{\sqrt{y^2-1}}\right),
\end{equation}
we find that the right hand side of \ref{K3lambdaF} satisfies the associated Legendre equation for degree $\nu+\lambda$ and order $\mu$ provided that the function
\begin{equation}
\label{Fnu+lambda_test}
u^{-\lambda}(1+u)^{-\lambda+\frac{1}{2}}\frac{d}{du}\left[\left( \frac{Y-1}{Y+1}\right)^{(\nu+\frac{1}{2})/2}(Y^2-1)^{-1/4}F_\nu^\mu\left( \frac{Y}{\sqrt{Y^2-1}}\right)\right]
\end{equation}
vanishes at the endpoints of the contour. This condition is satisfied for $F_\nu^\mu=P_\nu^\mu$ in \ref{K3lambdaF-1} provided ${\rm Re} (\nu+\lambda-\mu+1)>0$, and for $F_\nu^\mu=Q_\nu^\mu$ provided ${\rm Re} (\nu+\lambda\pm\mu+1)>0$.

The constant of proportionality $N'_+$ in \ref{K3lambdaF-1} can be determined by changing to $Y$ as the variable of integration in that equation, and then determining the asymptotic limits of the two sides for $y\rightarrow\infty$.   We find that 
\begin{equation}
\label{N'+}
N'_+(\nu,\mu,\lambda) = e^{-i\pi\lambda} \frac{\Gamma(\nu+\lambda-\mu+1)}{\Gamma(\nu-\mu+1)}
\end{equation}
for both $P_\nu^\mu$ and $Q_\nu^\mu$. The first can be established easily by identifying the characteristic power behavior $P_\nu^\mu(z)\propto [(z+1)/(z-1)]^{\mu/2}$ for $z\rightarrow 1$ on the two sides of the equation. The coefficient for $Q_\nu^\mu$ follows from the relation
\begin{equation}
\label{Q=P-P}
e^{-i\pi\mu}Q_\nu^\mu(z)=\frac{\pi}{2\sin{\pi\mu}}\left[P_\nu^\mu(z)- \frac{\Gamma(\nu+\mu+1)}{\Gamma(\nu-\mu+1)}P_\nu^{-\mu}(z)\right].
\end{equation}

Using the result for $N'_+$ in \ref{K3lambdaF-1}, we find that
\begin{eqnarray}
F_{\nu+\lambda}^\mu\left(\frac{y}{\sqrt{y^2-1}}\right) &=& \frac{1}{2\pi i}e^{i\pi\lambda}\frac{\Gamma(\lambda+1)\Gamma(\nu-\mu+1)}{\Gamma(\nu+\lambda -\mu+1} \nonumber \\
\label{K3lambdaF-2}
&& \times \int_{(\infty, 0+,\infty)}\frac{du}{u^{\lambda+1}} \left(\frac{y^2-1}{Y^2-1} \right)^{(\nu+1)/2} F_\nu^\mu\left(\frac{Y}{\sqrt{Y^2-1}}\right),
\end{eqnarray}
where ${\rm Re}(\nu+\lambda-\mu+1)>0$ for $P_\nu^\mu$, and ${\rm Re}(\nu+\lambda\pm\mu+1)>0$ for $Q_\nu^\mu$. Alternatively, taking $Y$ as the integration variable,
\begin{eqnarray}
F_{\nu+\lambda}^\mu\left(\frac{y}{\sqrt{y^2-1}} \right)  &=& \frac{1}{2\pi i}e^{i\pi\lambda} \frac{\Gamma(\lambda+1)\Gamma(\nu-\mu+1)}{\Gamma(\nu+\lambda -\mu+1)}(y^2-1)^{(\nu+\lambda+1)/2} \nonumber \\
\label{1st_transform}
&& \times\int_{(\infty,y+,\infty)}\frac{dY}{(Y-y)^{\lambda+1}}(Y^2-1)^{-(\nu + 1)/2}F_\nu^\mu\left(\frac{Y}{\sqrt{Y^2-1}}\right).
\end{eqnarray}
This expression has the form of a Weyl fractional integral \cite{TIT}, \S~13.2, but is apparently not known in the general case.

The substitutions $Y=\coth{\theta'}$, $y=\coth{\theta}$ in \ref{1st_transform} give the expression
\begin{eqnarray}
F_{\nu+\lambda}^\mu(\cosh{\theta}) &=& -\frac{1}{2\pi i} e^{i\pi\lambda} \frac{\Gamma(\lambda+1)\Gamma(\nu-\mu+1)}{\Gamma(\nu+\lambda -\mu+1} (\sinh{\theta})^{\nu+\lambda+1} \nonumber
\\
\label{2nd_transform}
&&\times \int_{(0,\theta+,0)}\frac{d\theta'}{\sinh^2\theta'}\frac{1}{(\coth{\theta'} - \coth{\theta})^{\lambda+1}}(\sinh{\theta'})^{\nu+1} F_\nu^\mu(\cosh{\theta'}) \nonumber
\\
&=&  -\frac{1}{2\pi i} e^{i\pi\lambda} \frac{\Gamma(\lambda+1)\Gamma(\nu-\mu+1)}{\Gamma(\nu+\lambda -\mu+1} \nonumber
\\
\label{2nd_transform_2}
&& \times \int_{(0,\theta+,0)}\frac{d\theta'}{\sinh\theta'} \left( \frac{\sinh{\theta'}}{\sinh{\theta}}\right)^\nu \left(\frac{\sinh{\theta'}}{\sinh{(\theta-\theta')}}\right)^{\lambda+1} F_\nu^\mu(\cosh{\theta'}),
\end{eqnarray}
while the substitutions $Y=\cosh{\phi'}$, $y=\cosh{\phi}$ give
\begin{eqnarray}
F_{\nu+\lambda}^\mu(\coth{\phi}) &=& \frac{1}{2\pi i} e^{i\pi\lambda} \frac{\Gamma(\lambda+1)\Gamma(\nu-\mu+1)}{\Gamma(\nu+\lambda -\mu+1} (\sinh{\phi})^{\nu+\lambda+1} \nonumber
\\
\label{3rd_transform}
&& \times \int_{(\infty,\phi+,\infty)}\frac{\sinh{\phi'}\,d\phi'}{(\cosh{\phi'} -\cosh{\phi})^{\lambda+1}}(\sinh{\phi'})^{-\nu-1}F_\nu^\mu(\coth{\phi'}).
\end{eqnarray}
Finally, with $y=z/\sqrt{z^2-1}$ and $Y=(z+u)/\sqrt{z^2-1}$ in \ref{K3lambdaF-2},
\begin{eqnarray}
F_{\nu+\lambda}^\mu(z) &=& \frac{1}{2\pi i} e^{i\pi\lambda} \frac{\Gamma(\lambda+1)\Gamma(\nu-\mu+1)}{\Gamma(\nu+\lambda -\mu+1} \nonumber
\\
\label{5th_transform}
&& \times \int_{(\infty,0+,\infty)} \frac{du}{u^{\lambda+1}}(u^2+2uz+1)^{-(\nu+1)/2}F_\nu^\mu\left(\frac{z+u}{\sqrt{u^2+2zu+1}}\right).
\end{eqnarray}

The foregoing relations for $K_3^\lambda$ are of the Weyl type. The natural endpoint for a Riemann-type contour for the original integrand in \ref{K3lambdaF} is at $Y=1$ or $xu=\sqrt{(y-1)/(y+1)}$. $x$ can again be scaled out, and the condition that the right hand side of \ref{K3lambdaF} define a solution of the associated Legendre equation reduces to the requirement that the function in \ref{Fnu+lambda_test}, with $(1+u)^{-\lambda+\frac{1}{2}}$ replaced by $(1-u)^{-\lambda+\frac{1}{2}}$ and $Y=y-u\sqrt{y^2-1}$, vanish at the endpoints. It fails to vanish for $F_\nu^\mu$ equal to either $P_\nu^\mu$ or $Q_\nu^\mu$, and there is no Riemann-type expression for $K_3^\lambda$.


\subsection{Relations for $P_3^\lambda$}
\label{subsec:P3^lambda}

The action of the operator $P_3^\lambda$ on $x^\nu F_\nu^\mu$ decreases $\nu$ by $\lambda$ as shown by the commutation relation \ref{DPlambda}. Thus, using the variable $y=z/\sqrt{z^2-1}$ and the expression in \ref{eP3action} for the action of $e^{-uP_3}$, we obtain the Weyl-type relation
\begin{eqnarray}
P_3^\lambda x^\nu F_\nu^\mu\left(\frac{y}{\sqrt{y^2-1}}\right) &=&
N'_-(\nu,\mu,\lambda)x^{\nu-\lambda}
\left(F'\right)_{\nu-\lambda}^\mu\left(\frac{y}{\sqrt{y^2-1}}\right) \nonumber \\
\label{P3lambdaF}
&& \hspace*{-30pt}=\frac{1}{2\pi i}e^{i\pi\lambda}\Gamma(\lambda+1) 
\int_{C_W}\frac{du}{u^{\lambda+1}}\left(\frac{Y^2-1}{y^2-1} \right)^{\nu/2} F_\nu^\mu\left(\frac{Y}{\sqrt{Y^2-1}}\right), 
\end{eqnarray}
where $Y=y-\frac{u}{x}\sqrt{y^2-1}$. We will again take the initial integration contour to run to $|u|\rightarrow\infty$ above the singularities of the integrand at $u/x=\sqrt{(y\pm 1)(y\mp 1)}$, scale out the dependence on $x$, rotate the contour counterclockwise by $\pi$, and replace $u$ by $e^{i\pi}u$ to reach the standard contour $(\infty,0+,\infty)$. This gives the expression
\begin{eqnarray}
N'_-(\nu,\mu,\lambda)\left(F'\right)_{\nu-\lambda}^\mu\left(\frac{y}{\sqrt{y^2-1}}\right) &=& \frac{1}{2\pi i}\Gamma(\lambda+1) \int_{(\infty,0+,\infty)}\frac{du}{u^{\lambda+1}} \nonumber
\\
\label{P3lambdaF-1}
&& \times  \left(\frac{Y^2-1}{y^2-1} \right)^{\nu/2} F_\nu^\mu\left(\frac{Y}{\sqrt{Y^2-1}}\right),
\end{eqnarray}
\addtocounter{footnote}{-1}
where $Y$ is now given by $Y=y+u\sqrt{y^2-1}$ and we have acquired an extra phase $e^{-i\pi\lambda}$. By considering the modified Legendre equation satisfied by the function\footnotemark
\begin{equation}
\label{vmunu}
\left(\frac{y+1}{y-1}\right)^{(\nu+\frac{1}{2})/2}(y^2-1)^{-1/4} F_\nu^\mu \left(\frac{y}{\sqrt{y^2-1}}\right),
\end{equation}
we find that the right hand side of \ref{P3lambdaF} satisfies the associated Legendre equation for degree $\nu-\lambda$ and order $\mu$ provided that the function
\begin{equation}
\label{Fnu-lambda_test}
u^{-\lambda}(1+u)^{\nu+\frac{3}{2}}\frac{d}{du}\left[\left( \frac{Y+1}{Y-1}\right)^{(\nu+\frac{1}{2})/2}(Y^2-1)^{-1/4}F_\nu^\mu\left( \frac{Y}{\sqrt{Y^2-1}}\right)\right]
\end{equation}
vanishes at the endpoints of the contour. This condition is satisfied for $F_\nu^\mu=P_\nu^\mu$ in \ref{P3lambdaF-1} provided ${\rm Re} (-\nu+\lambda-\mu)>0$. A comparison of the asymptotic limits of the two sides of \ref{P3lambdaF-1} for $y\rightarrow\infty$ gives
\begin{equation}
\label{N'-result}
N'_-(\nu,\mu,\lambda)= e^{-i\pi\lambda}\frac{\Gamma(-\nu+\lambda-\mu)}{\Gamma( -\nu-\mu)},
\end{equation}
and
\begin{eqnarray}
P_{\nu-\lambda}^\mu\left(\frac{y}{\sqrt{y^2-1}}\right) &=& \frac{1}{2\pi i} e^{i\pi\lambda}\Gamma(\lambda+1)\frac{\Gamma(- \nu-\mu)}{\Gamma(-\nu+\lambda-\mu)}  
\int_{(\infty,0+,\infty}\frac{du}{u^{\lambda+1}} \nonumber
\\
\label{P3lambdaF-2}
&& \times \left(\frac{Y^2-1}{y^2-1}\right)^{\nu/2} P_\nu^\mu\left(\frac{Y}{\sqrt{Y^2-1}}\right).
\end{eqnarray}

The corresponding result for $F_\nu^\mu=Q_\nu^\mu$ involves an extra term related through the Whipple transformation \cite{HTF} 3.3.2(13,14) to that found for $M_+^\lambda$ in \S\ref{subsec:M+Weyl}, and a different coefficient. We find that
\begin{eqnarray}
\frac{e^{i\pi\lambda}}{2\pi i}\Gamma(\lambda+1) \frac{\Gamma(\nu-\lambda+\mu+1)}{\Gamma(\nu+\mu+1)}\int_{(\infty,0+,\infty)} \frac{du}{u^{\lambda+1}} \left(\frac{Y^2-1}{y^2-1}\right)^{\nu/2}  e^{-i\pi\mu} Q_\nu^\mu\left(\frac{Y}{\sqrt{Y^2-1}}\right) \nonumber
\\
\label{P3lambdaQWeyl}
= e^{-i\pi\mu} Q_{\nu-\lambda}^\mu\left(\frac{y}{\sqrt{y^2-1}}\right) - \pi\frac{\cos{\pi\mu}\sin{\pi\lambda}}{\sin{\pi(\nu-\lambda+\mu)}} P_{\nu-\lambda}^\mu\left(\frac{y}{\sqrt{y^2-1}}\right), 
\end{eqnarray}
$Y=y+u\sqrt{y^2-1}$, ${\rm Re}(\lambda-\nu\pm\mu)>0$.

The condition for there to be a Riemann-type representation for $P_3^\lambda$ is given by \ref{Fnu-lambda_test} with $u+1$ replaced by $1-u$ and $Y=y-u(y-1)$. The result must vanish for $u=1$. This condition is satisfied for $F_\nu^\mu=Q_\nu^\mu$ for ${\rm Re}(\nu+\frac{3}{2})>0$, but is not satisfied for $P_\nu^\mu$. The Riemann integral for $Q_\nu^\mu$ is 
\begin{eqnarray}
Q_{\nu-\lambda}^\mu\left(\frac{y}{\sqrt{y^2-1}}\right) &=& \frac{1}{2\pi i} e^{i\pi\lambda}\Gamma(\lambda+1) \frac{\Gamma(\nu-\lambda+\mu+1}{\Gamma(\nu+\mu+1)}\int_{(u_0,0+,u_0)} \frac{du}{u^{\lambda+1}} \nonumber
\\
\label{P3_Riemann}
&& \times \left(\frac{Y^2-1}{y^2-1}\right)^{\nu/2}  Q_\nu^\mu\left(\frac{Y}{\sqrt{Y^2-1}}\right), 
\end{eqnarray}

where $Y=y-u\sqrt{y^2-1}$ and $u_0=\sqrt{(y-1)/(y+1)}$.


\section{Integral representations for associated Legendre functions}
\label{sec:LengendreIntegralReps}

\subsection{Representations using $M_\pm^\lambda$}
\label{subsubsec:IntRepsMpm}

The fractional-operator relators $M_\pm^\lambda t^\mu F_\nu^\mu(z)$ $= N_\pm t^{\mu\pm\lambda}F_\nu^{'\mu}(z)$ derived above can be converted into integral representations for the associated Legendre functions by a choice of $\mu$ for which the initial function is elementary. Thus, using $M_+^\lambda$, the choices
\begin{equation}
\label{Q^1/2(z)}
e^{-i\pi/2}\,Q_\nu^{1/2}(z) = \sqrt{\frac{\pi}{2}}(z^2-1)^{-1/4}\, [z+\sqrt{z^2-1}]^{-\nu-\frac{1}{2}}
\end{equation}
in \ref{Q-Wfrac} and
\begin{equation}
\label{Q^1/2(theta)}
e^{-i\pi/2}\,Q_\nu^{1/2}(\cosh{\theta}) = \sqrt{\frac{\pi}{2}}\sinh^{-1/2}{\theta}\, e^{-(\nu+\frac{1}{2})\theta}
\end{equation}
in \ref{Q-W_hyper} give simple Weyl-type integral representations for $Q_\nu^{\lambda+\frac{1}{2}}$. Choosing $\lambda=\mu-\frac{1}{2}$ in these expressions, we get, respectively,
\begin{eqnarray}
(z^2-1)^{-\mu/2}e^{-i\pi\mu}Q_\nu^\mu(z) &=& \frac{1}{2\pi i}e^{i\pi(\mu- \frac{1}{2})}\Gamma\left(\mu+\frac{1}{2}\right)\sqrt{\frac{\pi}{2}} \int_{(\infty, z+, \infty)}\frac{dZ}{(Z-z)^{\mu+\frac{1}{2}}} \nonumber
\\
\label{Qnumu_int_1}
&& \times (Z^2-1)^{-1/2} \left(Z+\sqrt{Z^2-1}\right)^{-\nu-\frac{1}{2}}
\end{eqnarray}
and
\begin{eqnarray}
e^{-i\pi\mu}Q_\nu^\mu(\cosh{\theta}) &=&\frac{1}{2\pi i}e^{i\pi(\mu- \frac{1}{2})}\Gamma\left(\mu+\frac{1}{2}\right)\sqrt{\frac{\pi}{2}} \sinh^\mu{\theta} \nonumber 
\\
\label{Qnumu_int_2}
&& \times \int_{(\infty, \theta+, \infty)} \frac{d\theta'}{(\cosh{\theta'}-\cosh{\theta} )^{\mu+\frac{1}{2}}}e^{-(\nu+\frac{1}{2})\theta'},
\end{eqnarray}
for ${\rm Re}(\nu+\mu+1)>0$. If ${\rm Re}\mu<\frac{1}{2}$, the contour integrals can be collapsed, and the resulting form for \ref{Qnumu_int_2} reduces to the standard integral representation \cite{HTF} 3.7(4). The fractional operator approach provides an interpretation of this result through its connection to the group SO(2,1).

The result in \ref{Qnumu_int_2} can be transformed further by distorting the integration contour  to $(\infty+i\pi,i\pi,-i\pi,\infty-i\pi)$, with $\theta'=0$ circumvented on the left. The result reduces for ${\rm Re}\,\mu<1/2$ to \cite{HTF} 3.7(10). Similar manipulations are possible for the following expressions. For collections of known representations, see \cite{HTF,whittaker,hobson}.

The use of \ref{Qnumu_int_2} and
\begin{equation}
\label{P^1/2(theta)}
P_\nu^{1/2}(\cosh{\theta}) = \sqrt{\frac{2}{\pi}}\sinh^{-1/2}{\theta}\,\cosh{\left(\nu+\frac{1}{2}\right) \theta}
\end{equation}
in the version of \ref{P-W} obtained by the substitutions $z=\cosh{\theta}$, $z+u\sqrt{z^2-1}=\cosh{\theta'}$ in the integral  on the first line gives a representation for $P_\nu^\mu(\cosh{\theta})$ analogous to \ref{Qnumu_int_2}.   

A second and more natural form of $M_+^\lambda P_\nu^\mu$ is given by the Riemann-type integral \ref{TIT13.1(54)}. Using $P_\nu^{1/2}$ as the input function, making the substitutions $z=\cosh{\theta}$, $Z'= \cosh{\theta'}$, and choosing $\lambda=\mu-\frac{1}{2}$, we obtain the integral representation
\begin{eqnarray}
P_\nu^\mu(\cosh{\theta}) &=& \frac{1}{2\pi i} \Gamma(\mu+\frac{1}{2}) \sqrt{\frac{2}{\pi}} \sinh^\mu{\theta} \nonumber
\\
\label{Pnumu_int_1}
&& \times \int_{(0,\theta+,0)}\frac{d\theta'}{(\cosh{\theta'}-\cosh{\theta})^{\mu +\frac{1}{2}}}\cosh{\left(\nu+\frac{1}{2}\right)\theta'},
\end{eqnarray}
where the denominator is now taken to have its principal phase, $-\pi\leq {\rm arg}(\cosh{\theta'}-\cosh{\theta})\leq \pi$, and $\nu$ and $\mu$ are arbitrary. If ${\rm Re}\mu<1/2$, the contour can be collapsed and \ref{Pnumu_int_1} reduces to \cite{HTF} 3.7(9).

We can get alternative integral representations for $Q_\nu^\mu$ and $P_\nu^\mu$ by using the relations 
\begin{eqnarray}
\label{Qnu^nu+1}
e^{-i\pi(\nu+1)}Q_\nu^{\nu+1}(z) &=& 2^\nu\Gamma(\nu+1) (z^2-1)^{-(\nu+1)/2},
\\
\label{Pnu^-nu}
P_\nu^{-\nu}(z) &=& \frac{2^{-\nu}}{\Gamma(\nu+1)} (z^2-1)^{\nu/2}.
\end{eqnarray}
Thus, from the Weyl-type integral \ref{Q-W}, 
\begin{eqnarray}
e^{-i\pi(\nu+\lambda+1)}Q_\nu^{\nu+\lambda+1}(z) &=& \frac{1}{2\pi i} e^{i\pi\lambda}2^\nu\Gamma(\nu+1)\Gamma(\lambda+1) (z^2-1)^{(\nu+1)/2} \nonumber
\\
\label{Qnumu_int_3}
&& \times \int_{(\infty,0+,\infty)}\frac{du}{u^{\lambda+1}}(Z^2-1)^{-\nu-1},
\end{eqnarray}
${\rm Re}(2\nu+\lambda+2)>0$, $Z=z+u\sqrt{z^2-1}$. The choice $\lambda=\mu-\nu-1$ then gives
\begin{eqnarray}
e^{-i\pi\mu}Q_\nu^\mu(z) &=& \frac{1}{2\pi i}e^{i\pi(\mu-\nu-1)} 2^\nu\Gamma(\nu+1)\Gamma(-\nu+\mu) (z^2-1)^{\mu/2} \nonumber
\\
\label{Qnumu_int_4}
&& \times \int_{(\infty,0+,\infty)}\frac{du}{u^{\mu-\nu}}(Z^2-1)^{-\nu-1}
\\
&=& 
\frac{1}{2\pi i}e^{i\pi(\mu-\nu-1)} 2^\nu\Gamma(\nu+1)\Gamma(-\nu+\mu) (z^2-1)^{\mu/2} \nonumber
\\
\label{Qnumu_int_5}
&& \times \int_{(\infty,z+,\infty)}\frac{dZ}{(Z-z)^{\mu-\nu}}(Z^2-1)^{-\nu-1}, \end{eqnarray}
${\rm Re}(\nu+\mu+1)>0$. The substitutions $z=\cosh{\theta}$, $Z=\cosh{\theta'}$ in the second form  gives an expression for $Q_\nu^\mu(\cosh{\theta})$ different from \ref{Qnumu_int_2},
\begin{eqnarray}
e^{-i\pi\mu}Q_\nu^\mu(\cosh{\theta}) &=& \frac{1}{2\pi i}e^{i\pi(\mu-\nu-1)} 2^\nu\Gamma(\nu+1)\Gamma(-\nu+\mu) \sinh^\mu{\theta} \nonumber
\\
\label{Qnumu_int_6}
&& \times \int_{(\infty,\theta_+,\infty)}\frac{d\theta'}{(\cosh{\theta'}-\cosh{\theta})^{\mu -\nu}}\sinh^{-2\nu-1}{\theta'},
\end{eqnarray}
${\rm Re}(\nu+\mu+1)>0$.

A similar construction starting from the Riemann-type integral \ref{TIT13.1(54)-2} for $M_+^\lambda$ with $\lambda=\nu+\mu$ and $P_\nu^{-\nu}$, \ref{Pnu^-nu}, as the input gives
\begin{eqnarray}
P_\nu^\mu(z) &=& \frac{1}{2\pi i}2^{-\nu} \frac{\Gamma(\nu+\mu+1)}{\Gamma(\nu+1)} (z^2-1)^{\mu/2} \nonumber
\\
\label{Pnumu_int_2}
&& \times \int_{(1,z+,1)}\frac{dZ}{(Z-z)^{\nu+\mu+1}}(Z^2-1)^\nu,
\end{eqnarray}
$|{\rm arg}(Z-z)|\leq\pi$, ${\rm Re}(\nu+1)>0$. With $z=\cosh{\theta}$, $Z=\cosh{\theta'}$, this becomes
\begin{eqnarray}
P_\nu^\mu(\cosh{\theta}) &=& \frac{1}{2\pi i} 2^{-\nu}\frac{\Gamma(\nu+\mu+1)}{\Gamma(\nu+1)}\sinh^\mu{\theta} \nonumber
\\
\label{Pnumu_int_3}
&& \times \int_{(0,\theta+,0)} \frac{d\theta'}{(\cosh{\theta'}-\cosh{\theta})^{\nu+\mu+1}}\sinh^{2\nu+1}{\theta'},
\end{eqnarray}
where $-\pi\leq{\rm arg}(\cosh{\theta'}-\cosh{\theta})\leq\pi$.

We can obtain further integral representations for $Q_\nu^\mu$ and $P_\nu^\mu$ using the Weyl-type integrals for $M_-^\lambda$ in \ref{Q-W3} and \ref{P-W3}, respectively. Thus, using \ref{Q-W3} with the input function $Q_\nu^{1/2}$, \ref{Q^1/2(z)} and $\lambda=-\mu+\frac{1}{2}$, we find that
\begin{eqnarray}
e^{-i\pi\mu}Q_\nu^\mu(z) &=& \frac{1}{2\pi i}e^{i\pi(-\mu+\frac{1}{2})} \frac{\Gamma (-\mu+\frac{1}{2})}{(\nu+\frac{1}{2})}\frac{\Gamma(\nu+\mu+1)}{ \Gamma(\nu-\mu+1)}\sqrt{\frac{\pi}{2}} \nonumber
\\
\label{Qnumu_int_7}
&&
\times (z^2-1)^{-\mu/2} \int_{(\infty,0+,\infty)} \frac{dZ}{(Z-z)^{-\mu+\frac{3}{2}}} \left(Z+\sqrt{Z^2-1}\right)^{-\nu-\frac{1}{2}}
\end{eqnarray}
By changing to the angular variables $z=\cosh{\theta}$, $Z=\cosh{\theta'}$ and integrating once by parts, this can be rewritten as 
\begin{eqnarray}
e^{-i\pi\mu}Q_\nu^\mu(\cosh{\theta}) &=& \frac{1}{2\pi i}e^{-i\pi(\mu+\frac{1}{2})}\frac{\Gamma(-\mu+\frac{1}{2}) \Gamma(\nu+\mu+1)}{ \Gamma(\nu-\mu+1)}\sqrt{\frac{\pi}{2}} \nonumber
\\
\label{Qnumu_int_8}
&& \times (\sinh{\theta})^{-\mu} \int_{(\infty,\theta+,\infty)} d\theta'\, (\cosh{\theta'}-\cosh{\theta})^{\mu-\frac{1}{2}} e^{-(\nu+\frac{1}{2})\theta'}
\\
&=& \frac{\Gamma(\nu+\mu+1)}{\Gamma(\mu+\frac{1}{2})\Gamma(\nu-\mu+1)} \sqrt{\frac{\pi}{2}} \nonumber
\\
\label{Qnumu_int_9}
&& \times (\sinh{\theta})^{-\mu} \int_\theta^\infty d\theta'\,(\cosh{\theta'} - \cosh{\theta})^{\mu-\frac{1}{2}} e^{-(\nu+\frac{1}{2})
\theta'}.
\end{eqnarray}
The first expression holds for ${\rm Re}(\nu-\mu+1)>0$, and the second with the additional constraint that ${\rm Re}\,\mu>-1/2$. The same results can be  obtained without the partial integration by starting with $Q_\nu^{-1/2}$ and $\lambda=-\mu-\frac{1}{2}$.

Note that the integral in the expression \ref{Qnumu_int_8} for $Q_\nu^\mu$ is the same as that in the expression for $Q_\nu^{-\mu}$ obtained from \ref{Qnumu_int_2} with the replacement $\mu\rightarrow -\mu$. Comparison of the two results shows that 
\begin{equation}
\label{Qsymmetry}
e^{-i\pi\mu}Q_\nu^\mu(z)=\frac{\Gamma(\nu+\mu+1)}{\Gamma(\nu-\mu+1)} e^{i\pi\mu}Q_\nu^{-\mu}(z)
\end{equation}
as expected.

A similar calculation using \ref{P-W3} with the input function $P_\nu^{1/2}$ and $\lambda=\mu+\frac{1}{2}$ gives the integral representation
\begin{eqnarray}
P_\nu^{-\mu}(\cosh{\theta}) &=& \frac{1}{2\pi i}e^{i\pi(\mu-\frac{1}{2})}\frac{\sqrt{2\pi}}{\cos{\pi\nu}}\frac{\Gamma(\mu+\frac{1}{2})}{\Gamma(-\nu+\mu) \Gamma(\nu+\mu+1)} \nonumber
\\
\label{Pnumu_int_4}
&& \times \sinh^\mu{\theta} \int_{(\infty,\theta+,\infty)}\frac{d\theta'}{( \cosh{\theta'}-\cosh{\theta})^{\mu+\frac{1}{2}}} \sinh{(\nu+\frac{1}{2})\theta'},
\end{eqnarray}
valid for ${\rm Re}(\mu\pm(\nu+\frac{1}{2})+\frac{1}{2})>0$.

In contrast to the case of $M_+^\lambda$, the input functions $Q_\nu^{\nu+1}$ and $P_\nu^{-\nu}$, \ref{Qnu^nu+1} and \ref{Pnu^-nu}, do not give useful results for $M_-^\lambda$ because of the appearance of infinite coefficients in \ref{Q-W3} and \ref{P-W3}. The problem arises from the recurrence relation connected with the action of $M_-$. The coefficient of $F_\nu^{\mu-1}$ in \ref{Pnumu_recurrence2} vanishes for $\mu=-\nu$ or $\mu=\nu+1$, and all information about $F_\nu^{\mu-1}$ is lost for those values of $\mu$.  

There are no Riemann-type relations or corresponding integral representations involving $M_-^\lambda$.


\subsection{Representations using $K_3^\lambda$ and $P_3^\lambda$}
\label{subsubsec:IntRepsPK}

The fractional operator relation $K_3^\lambda x^\nu F_\nu^\mu = N'_+x^{\nu+\lambda}F_{\nu+\lambda}^\mu$ gives integral representations for the associated Legendre functions when used with $P_0^\mu$ or $Q_0^\mu$ as the input function. Thus, using $P_0^\mu$, \ref{P0mu}, in \ref{1st_transform} and replacing $\lambda$ by $\nu$ in the result, we find that
\begin{eqnarray}
P_\nu^\mu\left(\frac{y}{\sqrt{y^2-1}}\right) &=& \frac{1}{2\pi i}e^{i\pi\nu} \frac{\Gamma(\nu+1)}{\Gamma(\nu-\mu+1)}(y^2-1)^{(\nu+1)/2} \nonumber
\\
\label{Pnumu_int_5}
&& \times \int_{(\infty,y+,\infty)} \frac{dY}{(Y-y)^{\nu+1}}\frac{1}{\sqrt{Y^2-1}}\,(Y+\sqrt{Y^2-1})^\mu
\end{eqnarray}
for ${\rm Re}(\nu-\mu+1)>0$. A change to the angular variables $y=\cosh{\phi}$, $Y=\cosh{\phi'}$ gives
\begin{eqnarray}
P_\nu^\mu(\coth{\phi}) &=& \frac{1}{2\pi i}e^{i\pi\nu} \frac{\Gamma(\nu+1)}{\Gamma(\nu-\mu+1)}(\sinh{\phi})^{\nu+1} \nonumber
\\
\label{Pnumu_int_6}
&& \times \int_{(\infty,\phi+,\infty)} \frac{d\phi'}{(\cosh{\phi'}-\cosh{\phi})^{\nu+1}} e^{\mu\phi'}.
\end{eqnarray}
Finally, using $P_0^\mu$, \ref{P0mu}, in \ref{5th_transform},
\begin{eqnarray}
P_\nu^\mu(z) &=& \frac{1}{2\pi i}e^{i\pi\nu} \frac{\Gamma(\nu+1)}{\Gamma(\nu-\mu+1)}(z^2-1)^{-\mu/2} \nonumber
\\
\label{Pnumu_int_7}
&& \times \int_{(\infty,0+,\infty)} \frac{du}{u^{\nu+1}}\frac{1 }{\sqrt{u^2+2zu+1}}(z+u+\sqrt{u^2 + 2zu +1})^\mu,
\end{eqnarray}
which reduces for $\mu=0$ to
\begin{equation}
\label{Pnu0_int_8}
P_\nu(z) = \frac{1}{2\pi i}e^{i\pi\nu} \int_{(\infty,0+,\infty)} \frac{du}{u^{\nu+1}}\frac{1 }{\sqrt{u^2+2zu+1}},
\end{equation}
${\rm Re}(\nu+1)>0$.

We get similar integral representations for $Q_\nu^\mu$ by using the expression \ref{Q0mu} in \ref{K3lambdaF-2}-\ref{5th_transform}. Thus, in terms of the angular variable $z=\coth{\phi}$,
\begin{eqnarray}
Q_\nu^\mu(\coth{\phi}) &=& \frac{1}{2\pi i}e^{i\pi\nu} \frac{\Gamma(\nu+1)}{\Gamma(\nu-\mu+1)} \frac{\pi}{\sin{\pi\mu}}(\sinh{\phi})^{\nu+1} \nonumber
\\
\label{Qnumu_int_10}
&& \times \int_{(\infty,\phi+,\infty)} \frac{d\phi'}{(\cosh{\phi'}-\cosh{\phi})^{\nu+1}} \sinh{\mu\phi'},
\end{eqnarray}
${\rm Re}(\nu\pm\mu+1)>0$.

The Weyl-type relation \ref{P3lambdaF-2} for $P_3^\lambda x^\nu P_\nu^\mu(z)$ with $\nu=0$ reduces after the use of the relation $P_0^\mu(\coth{\phi'})=e^{\mu\phi'}/\Gamma(1-\mu)$, the change of variables $y=\cosh{\phi}$, $Y=\cosh{\phi'}$, and a partial integration to
\begin{eqnarray}
P_{-\lambda}^\mu(\coth{\phi}) &=& \frac{1}{2\pi i}e^{i\pi(\lambda-1)} \frac{\Gamma(\lambda)}{\Gamma(\lambda-\mu)}\sinh^\lambda{\phi} \nonumber
\\
\label{P_{-lambda}^mu}
&& \times \int_{(\infty,\phi+,\infty)}\frac{d\phi'}{(\cosh{\phi'}-\cosh{\phi})^\lambda} e^{\mu\phi'},
\end{eqnarray}
${\rm Re}(\lambda-\mu)>0$. The right hand side of this expression is the same for the choice $\lambda=\nu+1$ as that in the expression \ref{Pnumu_int_6} for $P_\nu^\mu$, and we find that $P_{-\nu-1}^\mu(\coth{\phi})=P_\nu^\mu(\coth{\phi})$ as expected. This symmetry relation is a consequence of the relation
\begin{eqnarray}
[K_3^\nu,P_3^{\nu+1}]h_0^\mu &=& \left[N'_+(-\nu-1,\mu,\nu)N'_-(0,\mu,\nu+1) \right. \nonumber
\\
\label{P3K3comm}
&& \left. - N'_-(\nu,\mu,\nu+1)N'_+(0,\mu,\nu) \right]h_0^\mu = 0,
\end{eqnarray}
where the coefficients $N'_+$ and $N'_-$ given in \ref{N'+} and \ref{N'-result} follow from the stepping relations in so(2,1).  

The corresponding Weyl-type relation for $P_3^\lambda x^\nu Q_\nu^\mu(z)$ in \ref{P3lambdaQWeyl} involves an extra term, and will not be given. 

The Riemann-type relation for $P_3^\lambda x^\nu Q_\nu^\mu$ obtained by using \ref{Q0mu} in \ref{P3_Riemann} gives the further integral representation 
\begin{eqnarray}
Q_{-\lambda}^\mu(\coth{\phi}) &=& \frac{1}{2\pi i}\Gamma(\lambda) \frac{\Gamma(-\lambda+\mu+1)}{\Gamma(\mu)}\sinh^\lambda{\phi} \nonumber
\\
\label{Qnumu_int_11}
&& \times \int_{(0,\phi+,0)} \frac{d\phi'}{(\cosh{\phi'}-\cosh{\phi})^\lambda} \sinh{\mu\phi'},
\end{eqnarray}
where the phase of the denominator it to be taken between $-\pi$ and $\pi$. While the form of the integrands in \ref{Qnumu_int_10} and \ref{Qnumu_int_11} is the same for the choice $\lambda=\nu+1$ in the latter, the integration contours are different, and the difference between the $Q_\nu^\mu$ and $Q_{-\nu-1}^\mu$ involves an admixture of $P_\nu^\mu$.


\subsection{Double-integral representations}
\label{subsec:double_integrals}

Combinations of the foregoing results give a number of double-integral representations for the associated Legendre functions according to the fractional relations
\begin{eqnarray}
\label{MpmK3_prod}
M_\pm^{\lambda'} K_3^\lambda x^\nu t^\mu F_\nu^\mu(z) &=& N_\pm(\nu+\lambda,\mu, \lambda') N'_+(\nu,\mu,\lambda) x^{\nu+\lambda}t^{\mu\pm\lambda'} F_{\nu+\lambda}^{'\mu\pm\lambda'}(z),
\\
\label{MpmP3_prod}
M_\pm^{\lambda'} P_3^\lambda x^\nu t^\mu F_\nu^\mu(z) &=& N_\pm(\nu-\lambda,\mu, \lambda') N'_-(\nu,\mu,\lambda) x^{\nu-\lambda}t^{\mu\pm\lambda'} F_{\nu-\lambda}^{'\mu\pm\lambda'}(z),
\\
\label{K3Mpm_prod}
K_3^\lambda M_\pm^{\lambda'} x^\nu t^\mu F_\nu^\mu(z) &=& 
N'_+(\nu,\mu\pm\lambda',\lambda) N_\pm(\nu,\mu,\lambda') 
x^{\nu+\lambda}t^{\mu\pm\lambda'} F_{\nu+\lambda}^{'\mu\pm\lambda'}(z),
\\
\label{P3Mpm_prod}
P_3^\lambda M_\pm^{\lambda'} x^\nu t^\mu F_\nu^\mu(z) &=& 
N'_+(\nu,\mu\pm\lambda',\lambda) N_\pm(\nu,\mu,\lambda') 
x^{\nu-\lambda}t^{\mu\pm\lambda'} F_{\nu-\lambda}^{'\mu\pm\lambda'}(z),
\end{eqnarray}
and similar relations for other products of the operators $M_\pm^\lambda$, $K_3^\lambda$, and $P_3^\lambda$. The appropriate definitions of the operators and the coefficients depend on the input functions. We will give only a few examples.

We first obtain a double integral for $P_\nu^\mu$ using the operator $M_+^\mu K_3^\nu$. We start with \ref{MpmK3_prod} with $P_0^0=1$ as the input function. The action of $K_3^\nu$ gives $P_\nu^0$, \ref{Pnu0_int_8}, where we have suppressed the factor $x^\nu$ in \ref{MpmK3_prod}. Acting a second time with $M_+^\mu$ and suppressing the resulting factor $t^\mu$ gives, through \ref{TIT13.1(54)-2}, 
\begin{eqnarray}
P_\nu^\mu(z) &=& \frac{1}{(2\pi i)^2} e^{i\pi\nu}\Gamma(\mu+1)
(z^2-1)^{\mu/2} \nonumber
\\
\label{MpmK3P00}
&& \times \int_{(1,z+,1)}\frac{dZ}{(Z-z)^{\mu+1} }\int_{(\infty,0+,\infty)} \frac{du}{u^{\nu+1}}\frac{1}{\sqrt{u^2+2Zu+1}}
\\
&=& \frac{1}{(2\pi i)^2} e^{i\pi\nu}\Gamma(\mu+1) \left(\frac{z+1}{z-1}\right)^{\mu/2} \nonumber
\\
\label{MpmK3P00-2}
&& \times \int_{(0,1+,0)}\frac{dt}{(t-1)^{\mu+1}}\int_{(\infty,0+,\infty)} \frac{du}{u^{\nu+1}}\frac{1}{\sqrt{(u+1)^2+2tu(z-1)}},
\end{eqnarray}
${\rm Re}(\nu+1)>0$. The second form follows from the substitution $Z=1+(z-1)t$. 

Note that the dependence of $P_\nu^\mu$ on $\nu$ and $\mu$ is separated in the two integrals in \ref{MpmK3P00} and \ref{MpmK3P00-2}. This will also be true in the following examples.

A similar calculation starting with $Q_0^0(z)$ and using \ref{K3lambdaF-2} and \ref{Q-Wfrac} gives 
\begin{eqnarray}
Q_\nu^\mu(z) &=& \frac{1}{(2\pi i)^2} e^{i\pi(\nu+\mu)}\Gamma(\mu+1) (z^2-1)^{\mu/2}\int_{(\infty,z+,\infty)}\frac{dZ}{(Z-z)^{\mu+1}} \nonumber
\\
\label{MpnK3Q00}
&& \times  \int_{(\infty,0+, \infty)} \frac{du}{u^{\nu+1}} \frac{1}{\sqrt{u^2+2Zu+1}} \ln{\frac{u+Z + \sqrt{u^2+2Zu+1}}{\sqrt{Z^2-1}}},
\end{eqnarray}
${\rm Re}(\nu+1)>0$, ${\rm Re}(\nu+\mu+1)>0$. 

If we consider $M_+^\mu M_+^\nu P_\nu^{-\nu}(z)$, use 
\ref{Pnu^-nu} and \ref{TIT13.1(54)-2} to get an expression for $P_\nu^0$ as in \ref{Pnumu_int_2}, and then use \ref{TIT13.1(54)-2} again, we find that
\begin{eqnarray}
P_\nu^\mu(z) &=& \frac{1}{(2\pi i)^2} 2^{-\nu}\Gamma(\mu+1) (z^2-1)^{\mu/2} \nonumber
\\
\label{MMP}
&& \times \int_{(1,z+,1)}\frac{dZ'}{(Z'-z)^{\mu+1}} \int_{(1,Z'+,1)} \frac{dZ}{(Z-Z')^{\nu+1}} (Z^2-1)^\nu
\\
&=& \frac{1}{(2\pi i)^2} 2^{-\nu}\Gamma(\mu+1) \left(\frac{z+1}{z-1} \right)^{\mu/2} \nonumber
\\
\label{MMP-2}
&& \times \int_{(0,1+,0)} \frac{dt}{(t-1)^{\mu+1}} \int_{(0,1+,0)} \frac{du}{(u-1)^{\nu+1}}u^\nu \left(1+ tu\frac{z-1}{2}\right)^\nu,
\end{eqnarray}
${\rm Re}(\nu+1)>0$. Similarly, the expression for $M_+^\mu M_+^{-\nu-1} Q_\nu^{\nu+1}$ obtained by using \ref{Qnumu_int_5} and \ref{Q-Wfrac} reduces to
\begin{eqnarray}
e^{-i\pi\mu}Q_\nu^\mu(z) &=& \frac{1}{(2\pi i)^2} e^{i\pi(\mu-\nu)} 2^\nu \frac{\pi}{\sin{\pi\nu}} \Gamma(\mu+1) (z^2-1)^{\mu/2} \nonumber
\\
\label{MMQ}
&& \times \int_{(\infty,z+,\infty)} \frac{dZ}{(Z-z)^{\mu+1}} \int_{(\infty,Z+,\infty)} \frac{ dZ'}{(Z^{'2}-1)^{\nu+1}}(Z'-Z)^\nu
\\
&=& \frac{1}{2}\frac{1}{(2\pi i)^2} e^{i\pi(\mu-\nu)} \frac{\pi}{\sin{\pi\nu}} \Gamma(\mu+1) \left(\frac{z+1}{z-1}\right)^{\mu/2} \nonumber
\\
\label{MMQ-2}
&& \times \int_{(\infty,1+,\infty)} \frac{dt}{(t-1)^{\mu+1}} \int_{(\infty,1+,\infty)} \frac{du}{u^{\nu+1}}(u-1)^\nu \left(1+\frac{z-1}{2}tu\right)^{-\nu-1}
\\
&=& -\frac{1}{2}\frac{1}{(2\pi i)^2} e^{i\pi(\mu-\nu)} \frac{\pi}{\sin{\pi\nu}} \Gamma(\mu+1) \left(\frac{z+1}{z-1}\right)^{\mu/2} \left(\frac{2}{z-1}\right)^{\nu+1} \nonumber
\\
\label{MMQ-3}
&& \times \int_{(0,1+,0)} dt\, \frac{t^{\nu+\mu}}{(t-1)^{\mu+1}} \int_{(0,1+,0)} du \, u^\nu(u-1)^\nu \left(1+\frac{2}{z-1}tu\right)^{-\nu-1},
\end{eqnarray}
${\rm Re}(\nu+\mu+1)>0$. The second expression follows from the first through the substitutions $Z'=1+u(Z-1)$, $Z=1+t(z-1)$. The last expression then follows from the replacements $u\rightarrow 1/u$, $t\rightarrow 1/t$, with a change in the phases so that $|{\rm arg}(t-1)|,\, |{\rm arg}(u-1)|\leq\pi$ in the final result.

As a final example, we obtain a double integral for $Q_\nu^\mu$ using $K_3^\nu P_3^{\mu-1}$ and an input function $Q_{\mu-1}^\mu$, \ref{Qnu^nu+1}. Using \ref{P3_Riemann} and \ref{1st_transform}, we find after some changes of variable that
\begin{eqnarray}
e^{-i\pi\mu}Q_\nu^\mu\left(\frac{y}{\sqrt{y^2-1}}\right) &=& \frac{1}{(2\pi i)^2} e^{i\pi\nu} \sqrt{\pi} \frac{\pi}{\sin{\pi\nu}} \frac{\Gamma(\nu+1)\Gamma(\mu+1)}{\Gamma(\nu-\mu+1) \Gamma(\mu+\frac{1}{2}} 2^{-\nu-1} \left(\frac{2}{y-1}\right)^{-\nu} \nonumber
\\
\label{K3P3Q}
&& \times \int_{(\infty,1+,\infty)} \frac{dt}{(t-1)^{\nu+1}}\left(1+\frac{y-1}{2}t\right)^{-1/2}
\\
&& \times \int_{(0,1+,0)} \frac{du}{(u-1)^\mu} u^{\mu-\frac{1}{2}} \left(1+\frac{y-1}{2}tu\right)^{\mu-\frac{1}{2}}, \nonumber
\end{eqnarray}
${\rm Re}\,\mu>-\frac{1}{2}$.


\section{Remarks}
\label{sec:Wigner-Inonu}

The results obtained here demonstrate the utility of fractional operators in deriving relations for the associated Legendre functions $F_\nu^\mu$. Standard discussions \cite{vilenkin,miller1,talman} show that those functions give unitary representations of the Lie groups SO(2,1) or SO(3) for special values of $\nu$ and $\mu$, with the groups typically acting as symmetry groups in applications where Legendre functions appear naturally. The operators $M_\pm$ in the Lie algebra  of so(2,1) and $K_3$ and $P_3$ in its conformal extension generate changes in $\nu$ and $\mu$ in integer steps when acting on those representations. We have been concerned here only with realizations of the algebras through differential operators, and are able in that context to define fractional operators which change $\nu$ and $\mu$ by arbitrary amounts. Thus, general functions $P_\nu^\mu$ and $Q_\nu^\mu$ can be constructed as in \ref{MpmK3P00} and \ref{MpnK3Q00} starting with the trivial realizations $P_0^0=1$, $Q_0^0=\frac{1}{2}\ln{[(z+1)/(z-1)]}$ or from other special cases.

The results on Bessel functions derived in \cite{FracOpsBessel} using fractional operator methods can also be derived as limiting cases of results here. The connection occurs geometrically through a Wigner-In\"{o}n\"{u} contraction \cite{gilmore} in which we consider infinitesimal transformations in SO(2,1) near the apex of the hyperboloid $H^2$ at $z=\cosh{\theta}=1$.   These are equivalent for $\theta$ sufficiently small to E(2) transformations in the tangent plane to $H^2$ at $\theta=0$. If we scale the angles with $\theta\rightarrow\vartheta/\nu$ and consider the limit $\nu\rightarrow\infty$, the associated Legendre equation \ref{Legendre_eq}, reduces to the hyperbolic Bessel equation
\begin{equation}
\label{hyper_Bessel}
\left(\frac{d^2}{d\vartheta^2}+\frac{1}{\vartheta}\frac{d}{d\vartheta} -\frac{\mu^2}{\vartheta^2}-1 \right)Z_\mu(\vartheta)=0,
\end{equation}
and the Bessel functions appear as confluent limits of the Legendre functions,
\begin{equation}
\label{confluent_limits}
\quad I_\mu(\vartheta) = \lim_{\nu\rightarrow\infty}\nu^\mu P_\nu^{-\mu}\left(\cosh{\frac{\vartheta}{\nu}}\right), \quad K_\mu(\vartheta) = \lim_{\nu\rightarrow\infty}\nu^{-\mu} e^{-i\pi\mu} Q_\nu^\mu\left(\cosh{\frac{\vartheta}{\nu}}\right),
\end{equation}
\cite{HTF} 7.8(1,4). In the same limit, the so(2,1) operators $M_\pm$ become multiples of the e(2) stepping operators $P_\pm$ used in \cite{FracOpsBessel},
\begin{eqnarray}
\label{MtoP}
M_+&=&-e^{i\phi}(\partial_\theta + i\coth{\theta}\,\partial_\phi)\rightarrow \nu\left(-t\partial_\vartheta +\frac{t^2}{\vartheta} \partial_t\right)=\nu P_+,
\\
M_-&=&e^{-i\phi}(\partial_\theta-i\coth{\theta}\,\partial_\phi)\rightarrow \nu\left(\frac{1}{t}\partial_\theta+\frac{1}{\theta}\partial_t\right)=\nu P_-,
\end{eqnarray}
where $P=(P_1,P_2)$ is the translation operator in the plane. The generator $M_3$ of rotations around the axis of $H^2$ is the same as the generator $J_3$ of rotations in the tangent plane defined in \cite{FracOpsBessel}, $M_3=J_3$.    
Upon scaling the group parameter or hyperbolic angle $u$ in \ref{Mpm^lambda}, with $u\rightarrow u/\nu$, we find that the fractional operators $M_\pm^\lambda$ transform as
\begin{equation}
\label{Mlambda_to_Plambda}
 M_\pm^\lambda = \frac{1}{2\pi i}e^{i\pi\lambda} \nu^\lambda \int_{(\infty,0+,\infty} \frac{du}{u^{\lambda+1}}e^{-uM_\pm/\nu} \stackrel{\longrightarrow}{\scriptstyle\nu\rightarrow\infty}\nu^\lambda  P_\pm^\lambda.
\end{equation}
The extra factors of $\nu$ are absorbed in the conformal limit, and relations given here for Legendre functions become relations for Bessel functions when the limit exists. As an example, the relation $M_+^\lambda t^\mu e^{-i\pi\mu} Q_\nu^\mu(\cosh{\theta}) = t^{\mu+1}e^{-i\pi(\mu+\lambda} Q_\nu^{\mu+\lambda} (\cosh{\theta})$, which gives \ref{Q-W} for Legendre functions, becomes $P_+^\lambda t^\mu K_\mu(\vartheta)=t^{\mu+\lambda}K_{\mu+\lambda}(\vartheta)$ for $\theta= \vartheta/\nu$, $\nu\rightarrow\infty$, and gives the relation
\begin{eqnarray}
K_{\nu+\lambda}(\vartheta) &=&  \frac{1}{2\pi i}e^{i\pi\lambda}\Gamma(\lambda+1) \int_{(\infty,0+,\infty)}\frac{du}{u^{\lambda+1}}\left(\frac{\vartheta^2}{ \vartheta^2 + 2u\vartheta}\right)^{\mu/2} \nonumber 
\\
\label{Knu+lambda}
&& \times K_\mu(\sqrt{\vartheta^2+2u\vartheta}). 
\end{eqnarray}
This is equivalent to equation 3.44 of \cite{FracOpsBessel} or, for ${\rm Re}\,\lambda<0$, to the known fractional integral \cite{TIT}, 13.2(59).

The methods discussed here can clearly be generalized to other special functions and their associated groups \cite{vilenkin}, for example, the Gegenbauer,\footnote{This is where our development of the fractional operators actually started.} Hermite or parabolic cylinder, Whittaker, Laguerre, and Jacobi functions. There are also intriguing second-order differential operators which have the properties of stepping operators in Lie algebras \cite{miller2,koornwinder,wolf1,wolf2,durand77} and can be exponentiated and integrated formally to obtain fractional operators, but the explicit action of the exponentials is not known in general.

\quad


\medskip
{\bf Acknowledgement:} The author would like to thank the faculty of the Institute for Advanced Study for their hospitality during the fall term of 1975 when the initial stages of this work were carried out, and the Aspen Center for Physics for its hospitality while parts of the final work were done.

\bibliographystyle{unsrt}
\bibliography{math}

\end{document}